\documentclass[prc, twocolumn, nofootinbib]{revtex4-1}
\usepackage{amsmath}
\usepackage{amsfonts}
\usepackage{amssymb}
\usepackage{graphicx}
\usepackage{color}
\usepackage{hyperref}

\usepackage{multirow}

\usepackage{lineno}
\newcommand{\result}[1]{#1}

\newcommand{\externalresult}[1]{#1}

\newcommand{\rhonuc}{\ensuremath{n_\mathrm{sat}}}

\newcommand{\PSRGWMmaxAgn}{\externalresult{\ensuremath{2.20^{+0.24}_{-0.18}}}} % 2.202458^{+0.238210}_{-0.184304}
\newcommand{\PSRGWRAgn}{\externalresult{\ensuremath{10.95^{+2.00}_{-1.37}}}} % 10.950000^{+2.000000}_{-1.365000}
\newcommand{\PSRGWLAgn}{\externalresult{\ensuremath{228^{+319}_{-134}}}} % 228.122963^{+319.377037}_{-133.782963}
\newcommand{\PSRGWPoneAgn}{\externalresult{\ensuremath{1.38^{+2.72}_{-1.31}}}} % 1.38254380993 2.71633826069 -1.30712695034 % 2.215077e+33^{+4.352049e+33}_{-2.094246e+33}

\newcommand{\PSRGWPtwoAgn}{\externalresult{\ensuremath{1.10^{+1.88}_{-1.09}}}} % 11.0278034712 18.826596191 -10.9474252306 % 1.766847e+34^{+3.016350e+34}_{-1.753969e+34}
\newcommand{\PSRGWPthreeAgn}{\externalresult{\ensuremath{0.71^{+0.60}_{-0.56}}}} % 70.624873421 60.4744897583 -55.5542654021 % 1.131534e+35^{+9.689071e+34}_{-8.900765e+34}
\newcommand{\PSRGWPfourAgn}{\externalresult{\ensuremath{2.01^{+0.95}_{-0.84}}}} % 200.785381552 95.3717345961 -83.6335030718 % 3.216933e+35^{+1.528022e+35}_{-1.339955e+35}
\newcommand{\PSRGWPsixAgn}{\externalresult{\ensuremath{5.71^{+2.57}_{-2.48}}}} % 570.70323349 256.838250444 -248.146752443 % 9.143664e+35^{+4.114998e+35}_{-3.975745e+35}

\newcommand{\PSRGWNICERMmaxAgn}{\externalresult{\ensuremath{2.22^{+0.30}_{-0.20}}}} % 2.224168^{+0.298808}_{-0.203338}
\newcommand{\PSRGWNICERRAgn}{\externalresult{\ensuremath{12.32^{+1.09}_{-1.47}}}} % 12.320000^{+1.090000}_{-1.470000}
\newcommand{\PSRGWNICERLAgn}{\externalresult{\ensuremath{451^{+241}_{-279}}}} % 450.900000^{+241.400000}_{-279.300000}
\newcommand{\PSRGWNICERPoneAgn}{\externalresult{\ensuremath{2.68^{+2.37}_{-2.48}}}} % 2.68137953068 2.37347869047 -2.47664345687 % 4.296039e+33^{+3.802728e+33}_{-3.968016e+33}

\newcommand{\PSRGWNICERPtwoAgn}{\externalresult{\ensuremath{2.38^{+1.66}_{-1.83}}}} % 23.808773833 16.6206447778 -18.2798394067 % 3.814582e+34^{+2.662918e+34}_{-2.928750e+34}
\newcommand{\PSRGWNICERPthreeAgn}{\externalresult{\ensuremath{0.98^{+0.59}_{-0.56}}}} % 97.969016567 59.2558025903 -55.7612590322 % 1.569635e+35^{+9.493816e+34}_{-8.933929e+34}
\newcommand{\PSRGWNICERPfourAgn}{\externalresult{\ensuremath{2.11^{+1.11}_{-0.73}}}} % 211.27306297 111.476967482 -73.3465488018 % 3.384964e+35^{+1.786056e+35}_{-1.175140e+35}
\newcommand{\PSRGWNICERPsixAgn}{\externalresult{\ensuremath{5.38^{+3.30}_{-2.66}}}} % 537.948008219 329.784468422 -266.105341032 % 8.618868e+35^{+5.283724e+35}_{-4.263473e+35}

\newcommand{\PSRGWMmaxOne}{\result{\ensuremath{2.20^{+0.25}_{-0.20}}}} % 2.202138^{+0.254502}_{-0.198693}
\newcommand{\PSRGWROne}{\result{\ensuremath{11.59^{+1.23}_{-1.19}}}} % 11.590000^{+1.230000}_{-1.190000}
\newcommand{\PSRGWLOne}{\result{\ensuremath{291^{+264}_{-175}}}} % 291.341571^{+263.958429}_{-174.641571}
\newcommand{\PSRGWPoneOne}{\result{\ensuremath{2.13^{+0.49}_{-0.46}}}} % 2.132314e+00^{+4.933101e-01}_{-4.585381e-01}
\newcommand{\PSRGWPtwoOne}{\result{\ensuremath{1.62^{+1.94}_{-0.99}}}} % 1.623404e+01^{+1.935459e+01}_{-9.887592e+00}
\newcommand{\PSRGWPthreeOne}{\result{\ensuremath{0.86^{+0.61}_{-0.57}}}} % 8.648190e+01^{+6.057668e+01}_{-5.694092e+01}
\newcommand{\PSRGWPfourOne}{\result{\ensuremath{2.03^{+0.90}_{-0.74}}}} % 2.032689e+02^{+9.015524e+01}_{-7.401141e+01}
\newcommand{\PSRGWPsixOne}{\result{\ensuremath{5.35^{+2.57}_{-2.78}}}} % 5.352266e+02^{+2.569203e+02}_{-2.781492e+02}

\newcommand{\PSRGWNICERMmaxOne}{\result{\ensuremath{2.25^{+0.32}_{-0.25}}}} % 2.254261^{+0.317660}_{-0.251523}
\newcommand{\PSRGWNICERROne}{\result{\ensuremath{12.59^{+0.64}_{-0.59}}}} % 12.590000^{+0.640000}_{-0.590000}
\newcommand{\PSRGWNICERLOne}{\result{\ensuremath{518^{+208}_{-163}}}} % 518.243176^{+208.056824}_{-162.543176}
\newcommand{\PSRGWNICERPoneOne}{\result{\ensuremath{2.27^{+0.58}_{-0.46}}}} % 2.269179e+00^{+5.811129e-01}_{-4.596209e-01}
\newcommand{\PSRGWNICERPtwoOne}{\result{\ensuremath{3.12^{+2.03}_{-1.42}}}} % 3.117322e+01^{+2.029404e+01}_{-1.422308e+01}
\newcommand{\PSRGWNICERPthreeOne}{\result{\ensuremath{1.12^{+0.59}_{-0.40}}}} % 1.121284e+02^{+5.910394e+01}_{-4.013801e+01}
\newcommand{\PSRGWNICERPfourOne}{\result{\ensuremath{2.12^{+1.07}_{-0.88}}}} % 2.117976e+02^{+1.074826e+02}_{-8.797738e+01}
\newcommand{\PSRGWNICERPsixOne}{\result{\ensuremath{4.95^{+3.05}_{-3.18}}}} % 4.954305e+02^{+3.054872e+02}_{-3.176852e+02}

\newcommand{\PSRGWMmaxTwo}{\result{\ensuremath{2.18^{+0.20}_{-0.17}}}} % 2.175922^{+0.197944}_{-0.171106}
\newcommand{\PSRGWRTwo}{\result{\ensuremath{11.21^{+1.23}_{-0.91}}}} % 11.210000^{+1.230000}_{-0.910000}
\newcommand{\PSRGWLTwo}{\result{\ensuremath{227^{+218}_{-108}}}} % 226.800000^{+218.000000}_{-107.900000}
\newcommand{\PSRGWPoneTwo}{\result{\ensuremath{2.19^{+0.49}_{-0.47}}}} % 2.187470e+00^{+4.935955e-01}_{-4.708841e-01}
\newcommand{\PSRGWPtwoTwo}{\result{\ensuremath{1.33^{+0.99}_{-0.51}}}} % 1.330293e+01^{+9.910899e+00}_{-5.110770e+00}
\newcommand{\PSRGWPthreeTwo}{\result{\ensuremath{0.68^{+0.51}_{-0.45}}}} % 6.827287e+01^{+5.071723e+01}_{-4.501739e+01}
\newcommand{\PSRGWPfourTwo}{\result{\ensuremath{2.00^{+0.78}_{-0.71}}}} % 1.995478e+02^{+7.841992e+01}_{-7.144722e+01}
\newcommand{\PSRGWPsixTwo}{\result{\ensuremath{5.41^{+2.20}_{-2.15}}}} % 5.414799e+02^{+2.200782e+02}_{-2.150231e+02}

\newcommand{\PSRGWNICERMmaxTwo}{\result{\ensuremath{2.21^{+0.28}_{-0.22}}}} % 2.210843^{+0.281606}_{-0.218868}
\newcommand{\PSRGWNICERRTwo}{\result{\ensuremath{12.43^{+0.53}_{-0.77}}}} % 12.429962^{+0.530038}_{-0.769962}
\newcommand{\PSRGWNICERLTwo}{\result{\ensuremath{465^{+125}_{-177}}}} % 465.000000^{+124.800000}_{-177.100000}
\newcommand{\PSRGWNICERPoneTwo}{\result{\ensuremath{2.54^{+0.60}_{-0.49}}}} % 2.543266e+00^{+5.961710e-01}_{-4.922999e-01}
\newcommand{\PSRGWNICERPtwoTwo}{\result{\ensuremath{2.61^{+0.94}_{-1.23}}}} % 2.607274e+01^{+9.438935e+00}_{-1.227737e+01}
\newcommand{\PSRGWNICERPthreeTwo}{\result{\ensuremath{1.05^{+0.49}_{-0.37}}}} % 1.050080e+02^{+4.862417e+01}_{-3.659611e+01}
\newcommand{\PSRGWNICERPfourTwo}{\result{\ensuremath{2.08^{+0.92}_{-0.64}}}} % 2.079696e+02^{+9.153552e+01}_{-6.389717e+01}
\newcommand{\PSRGWNICERPsixTwo}{\result{\ensuremath{4.99^{+2.86}_{-2.58}}}} % 4.989659e+02^{+2.862389e+02}_{-2.582840e+02}

\newcommand{\PSRGWMmaxMrg}{\result{\ensuremath{2.19^{+0.24}_{-0.18}}}} % 2.188540^{+0.235820}_{-0.182806}
\newcommand{\PSRGWRMrg}{\result{\ensuremath{11.40^{+1.38}_{-1.04}}}} % 11.400000^{+1.380000}_{-1.040000}
\newcommand{\PSRGWLMrg}{\result{\ensuremath{260^{+270}_{-140}}}} % 260.300000^{+270.000000}_{-140.000000}
\newcommand{\PSRGWPoneMrg}{\result{\ensuremath{2.15^{+0.64}_{-0.53}}}} % 2.146528e+00^{+6.411325e-01}_{-5.296601e-01}
\newcommand{\PSRGWPtwoMrg}{\result{\ensuremath{1.42^{+1.81}_{-0.84}}}} % 1.419387e+01^{+1.805017e+01}_{-8.447634e+00}

\newcommand{\PSRGWPtwoMrgINLINE}{\result{\ensuremath{14.2^{+18.1}_{-8.4}}}} % 1.419387e+01^{+1.805017e+01}_{-8.447634e+00}

\newcommand{\PSRGWPthreeMrg}{\result{\ensuremath{0.79^{+0.56}_{-0.55}}}} % 7.920124e+01^{+5.578025e+01}_{-5.457431e+01}
\newcommand{\PSRGWPfourMrg}{\result{\ensuremath{2.02^{+0.87}_{-0.73}}}} % 2.022429e+02^{+8.730556e+01}_{-7.288885e+01}
\newcommand{\PSRGWPsixMrg}{\result{\ensuremath{5.36^{+2.39}_{-2.53}}}} % 5.364293e+02^{+2.388892e+02}_{-2.532360e+02}

\newcommand{\PSRGWNICERMmaxMrg}{\result{\ensuremath{2.24^{+0.31}_{-0.23}}}} % 2.236112^{+0.305028}_{-0.232311}
\newcommand{\PSRGWNICERRMrg}{\result{\ensuremath{12.54^{+0.71}_{-0.63}}}} % 12.540000^{+0.710000}_{-0.630000}
\newcommand{\PSRGWNICERLMrg}{\result{\ensuremath{494^{+201}_{-166}}}} % 494.100000^{+200.900000}_{-166.400000}
\newcommand{\PSRGWNICERPoneMrg}{\result{\ensuremath{2.42^{+0.75}_{-0.66}}}} % 2.423888e+00^{+7.514007e-01}_{-6.633352e-01}
\newcommand{\PSRGWNICERPtwoMrg}{\result{\ensuremath{2.87^{+1.53}_{-1.50}}}} % 2.869825e+01^{+1.526091e+01}_{-1.502871e+01}

\newcommand{\PSRGWNICERPtwoMrgINLINE}{\result{\ensuremath{28.7^{+15.3}_{-15.0}}}} % 2.869825e+01^{+1.526091e+01}_{-1.502871e+01}

\newcommand{\PSRGWNICERPthreeMrg}{\result{\ensuremath{1.08^{+0.56}_{-0.38}}}} % 1.083437e+02^{+5.579982e+01}_{-3.803260e+01}
\newcommand{\PSRGWNICERPfourMrg}{\result{\ensuremath{2.11^{+1.08}_{-0.69}}}} % 2.107861e+02^{+1.077275e+02}_{-6.892369e+01}
\newcommand{\PSRGWNICERPfourMrgMeV}{\result{\ensuremath{211^{+108}_{-69}}}} % 2.107861e+02^{+1.077275e+02}_{-6.892369e+01}
\newcommand{\PSRGWNICERPsixMrg}{\result{\ensuremath{4.97^{+2.96}_{-2.98}}}} % 4.974155e+02^{+2.958218e+02}_{-2.982812e+02}

\newcommand{\EOS}{EoS}
\newcommand{\EOSs}{EoSs}
\newcommand{\EFT}{\ensuremath{\chi\text{EFT}}}
\newcommand{\NS}{NS}
\newcommand{\NSs}{NSs}
\newcommand{\QCD}{QCD}

\newcommand{\QMCbasic}{QMC}
\newcommand{\QMC}{QMC$_{\text{N}^2\text{LO,l}}^{(2018)}$}
\newcommand{\MBPTbasic}{MBPT}
\newcommand{\MBPT}{MBPT$_{\text{N}^3\text{LO,nl}}^{(2013)}$}

\newcommand{\GW}{GW}
\newcommand{\GWs}{GWs}
\newcommand{\PSR}{PSR}
\newcommand{\PSRs}{PSRs}
\newcommand{\NICER}{NICER}
\newcommand{\pmax}{\ensuremath{p_\mathrm{max}}}
\newcommand{\Bayes}{\ensuremath{B^\mathrm{theory}_\mathrm{agnostic}}}

\begin{document}

\title{
Direct Astrophysical Tests of Chiral Effective Field Theory at Supranuclear Densities
}

\author{Reed Essick}
    \email[Email: ]{reed.essick@gmail.com}
    \affiliation{Kavli  Institute  for  Cosmological  Physics,  The  University  of  Chicago, Chicago,  IL  60637, USA}

\author{Ingo Tews}
    \affiliation{Theoretical Division, Los Alamos National Laboratory, Los Alamos, NM 87545, USA}

\author{Philippe Landry}
    \affiliation{Gravitational-Wave Physics \& Astronomy Center, California State University, Fullerton, 800 N State College Blvd, Fullerton, CA 92831}

\author{Sanjay Reddy}
    \affiliation{Institute for Nuclear Theory, University of Washington, Seattle, WA 98195, USA}

\author{Daniel E. Holz}
    \affiliation{Kavli  Institute  for  Cosmological  Physics,  The  University  of  Chicago, Chicago,  IL  60637, USA}
    \affiliation{Enrico  Fermi  Institute, Department of Physics, and Department of Astronomy \& Astrophysics, The  University  of  Chicago, Chicago,  IL  60637, USA}

\date{\today} 

\begin{abstract}
Recent observations of neutron stars with gravitational waves and X-ray timing provide unprecedented access to the equation of state (\EOS) of cold dense matter at densities difficult to realize in terrestrial experiments.
At the same time, predictions for the \EOS~equipped with reliable uncertainty estimates from chiral effective field theory (\EFT) allow us to bound our theoretical ignorance.
In this work, we analyze astrophysical data using a nonparametric representation of the neutron-star \EOS~conditioned on \EFT~to directly constrain the underlying physical properties of the compact objects without introducing modeling systematics.
We discuss how the data alone constrain the \EOS~at high densities when we condition on \EFT~at low densities.
We also demonstrate how to exploit astrophysical data to directly test the predictions of \EFT~for the \EOS~up to twice nuclear saturation density, in order to estimate the density at which these predictions might break down.
We find that the existence of massive pulsars, gravitational waves from GW170817, and \NICER~observations of \PSR~J0030+0451 favor \EFT~predictions for the \EOS~up to nuclear saturation density over a more agnostic analysis by as much as a factor of \result{7} for the quantum Monte Carlo (\QMCbasic) calculations used in this work.
While \EFT~predictions using \QMCbasic~are fully consistent with gravitational-wave data up to twice nuclear saturation density, \NICER~observations suggest that the \EOS~stiffens relative to these predictions at or slightly above nuclear saturation density.
Additionally, for these \QMCbasic~calculations, we marginalize over the uncertainty in the density at which \EFT~begins to break down, constraining the radius of a $1.4\,M_\odot$ neutron star to $R_{1.4} = \PSRGWRMrg$ (\PSRGWNICERRMrg) km and the pressure at twice nuclear saturation density to $p(2\rhonuc) = \PSRGWPtwoMrgINLINE$ (\PSRGWNICERPtwoMrgINLINE) $\mathrm{MeV}/\mathrm{fm}^3$ with massive pulsar and gravitational-wave (and \NICER) data.
\end{abstract}

\maketitle

% report number: LA-UR-20-22615

%------------------------------------------------------------
\section{Introduction}
\label{sec:Intro}

The properties of dense, strongly interacting matter, described by the quantum chromodynamics (\QCD) phase diagram, remain elusive both experimentally and theoretically.
While recent progress on both fronts provides glimpses of the underlying physical interactions, large parts of the phase diagram are still unknown.

Experiments that collide heavy ions have helped constrain the properties of dense matter at finite temperature and small isospin asymmetry up to densities of 2--4$\rhonuc$~\cite{Danielewicz:2002pu}, where the nuclear saturation density $\rhonuc = 0.16$~nucleons/fm$^{3}$ is the baryon density encountered in the center of atomic nuclei.
Nuclear structure experiments, especially those pertaining to neutron-rich nuclei, continue to provide useful information about the properties of cold neutron-rich matter at sub-nuclear densities~\cite{Tsang:2012se, Roca-Maza:2015eza}.
In the cosmos, dense strongly-interacting matter can be found in the cores of neutron stars (\NSs), remnants of core-collapse supernovae that harbor matter with large isospin asymmetry at densities up to 10$\rhonuc$.
Theoretically, the description of the dense \QCD~matter encountered in \NSs~is feasible only at either low baryon density ($\lesssim 2 \rhonuc$) where strong interactions among nucleons are tractable~\cite{Hebeler:2009iv,Hebeler:2010jx} or at very large density ($\gtrsim 50 \rhonuc$) when interactions between quarks are weak~\cite{Kurkela:2009gj}.
At the intermediate densities relevant for the core of \NSs, no microscopic calculations with robust uncertainties exist so far (but see Ref.~\cite{Leonhardt:2019fua} for advances in symmetric matter).

The dense-matter equation of state (\EOS) relates the pressure $p$ and the energy density $\epsilon$ at a given baryon density $n$ and temperature $T$.
Of particular interest for \NSs~is the \EOS~of dense matter in the limit  $T \to 0$, because thermal energies in the interior of \NSs~are typically much smaller than the Fermi energy.
In this limit, the stars are barotropic, i.e., their density is a function of only their pressure.
Additionally, the relations between a \NS's mass and radius, compactness (mass/radius), and tidal response are uniquely determined by the \EOS.
Thus, observational constraints on \NS~masses, radii, compactnesses, or tidal responses can provide important insight about the low-temperature, high-density \QCD~phase diagram.

Over the past decade, radio and X-ray observations of \NSs~have measured \NS~masses and constrained their radii (for a recent review see Ref.~\cite{Ozel:2016oaf}).
Most notably, radio observations provided evidence for pulsars (\PSRs) with masses $\gtrsim2$ M$_\odot$~\cite{Demorest:2010bx,Antoniadis:2013pzd,Cromartie:2019kug}.
The most recent X-ray timing observations of \PSR~J0030+0451 by NASA's Neutron-Star Interior Composition Explorer (\NICER) mission~\cite{Riley_2019, Raaijmakers_2019, Miller:2019cac} were primarily sensitive to the \NS~compactness, although the radius might be more tightly constrained for other \PSRs. %as light emitted from hotspots on the \NS's surface experience more pronounced relativistic effects with increasing compactness and modify the temporal features of the observed X-ray pulse profile.
The detection of gravitational waves (\GWs) from binary neutron-star mergers by the Advanced LIGO~\cite{TheLIGOScientific:2014jea} and Virgo~\cite{TheVirgo:2014hva} interferometers, including GW170817~\cite{TheLIGOScientific:2017qsa} and GW190425~\cite{Abbott:2020uma}, simultaneously measured the components' masses and their tidal deformabilities, although the observations primarily constrain specific combinations of the masses and tidal deformabilities (the chirp mass and the binary tidal deformability) as they produce the largest observable effects in the \GW~signal.
The tidal deformabilities relate a \NS's quadrupole deformation to an external tidal field, like the one exerted by a binary companion.
As \NSs~deform in the tidal field induced by their companion, the deformation modifies the orbital phase, which in turn is directly imprinted in the observed \GW~strain.
A flurry of recent articles have studied in some detail how these observations, both individually and taken together, constrain the \EOS~and \NS~properties (see, e.g., \cite{Annala:2017llu,Abbott:2018exr,Most:2018hfd,Tews:2018chv,Greif:2018njt,Capano:2019eae,Miller:2019cac,Raaijmakers:2019dks,Dietrich:2020lps, Landry:2020}).
Although other macroscopic properties of \NSs~may also be observable, such as the moment of inertia~\cite{LyneBurgay2004, LattimerSchutz2005} and the maximum attainable spin frequency~\cite{Hessels2006}, our current knowledge of the \EOS~at densities above $\rhonuc$ is dominated by observations of massive \PSRs, \GWs, and X-ray timing.

Theoretical advances in recent years, largely achieved by employing nuclear interactions from chiral effective field theory (\EFT) and the development of computational methods to solve the nuclear many-body problem, provide theoretical constraints on the \EOS~of neutron matter up to 1--2 $\rhonuc$ with theoretical uncertainty estimates~\cite{Hebeler:2009iv,Tews:2012fj, Lynn:2015jua, Drischler:2016djf,Drischler:2020hwi}.
This means that, instead of providing a single estimate for the \EOS, \EFT~calculations provide an uncertainty range for the pressure at a given density and, hence, a prior process for the \EOS~at densities relevant for \NS~cores.
Several previous studies have used \EFT~constraints with theoretical uncertainties as priors on the \EOS~by assuming the validity of the theory up to a particular density~\citep{Tews:2012fj, Hebeler:2013nza, Annala:2017llu, Most:2018hfd, Lim:2018bkq, Tews:2018chv, Greif:2018njt, Capano:2019eae, Raaijmakers:2019dks,Dietrich:2020lps}.
We, instead, directly test for the breakdown of \EFT~predictions with astrophysical data and do not assert that the theory is correct up to even $\rhonuc/2$~\textit{a priori}.
To this effect, we condition on the entire theoretical prior process to construct a set of nested models, which faithfully reproduce \EFT~predictions up to higher and higher densities, showing how comparisons between theoretical predictions and astrophysical data can directly constrain the range of validity of a particular theory, in our case, \EFT.

Specifically, we make use of the \EFT~calculations of Ref.~\cite{Tews:2018kmu} using quantum Monte Carlo (\QMCbasic) methods~\cite{Carlson:2015} and local chiral interactions from Refs.~\cite{Gezerlis:2013ipa,Lynn:2015jua} to investigate the evidence for a possible breakdown of our \EFT~description between 1--2 $\rhonuc$.
We compare our findings with the neutron-matter calculations of Ref.~\cite{Tews:2012fj} using many-body perturbation theory (\MBPTbasic) and nonlocal \EFT~interactions from Refs.~\cite{Entem:2003ft,Epelbaum:2004fk} for reference.
We show that astrophysical data sets, by themselves, prefer \EFT~calculations over more agnostic priors, instead of forcing analysts to assert precise knowledge of the \EFT~breakdown density \textit{a priori}.
Furthermore, by marginalizing over the maximum density up to which \EFT~is valid, we allow the data to determine where we trust theoretical predictions.

\begin{figure*}
    \includegraphics[width=0.49\textwidth, clip=True, trim=0.3cm 0.3cm 0.3cm 0.4cm]{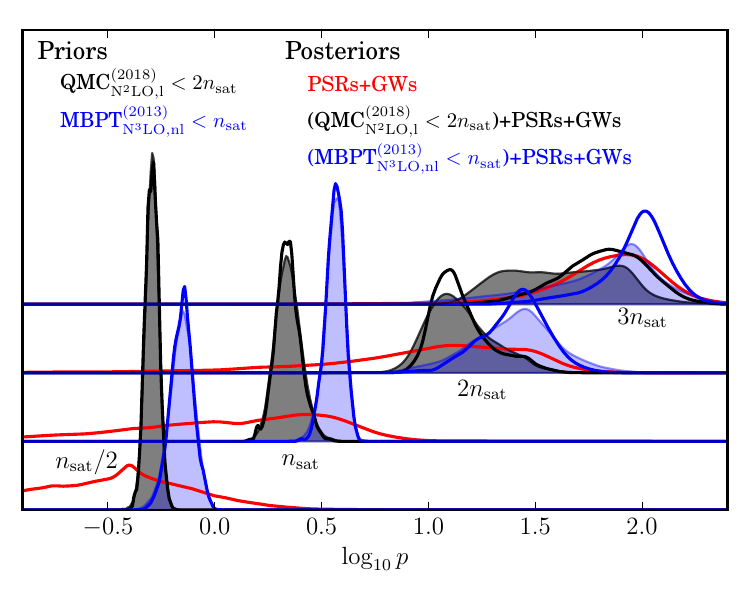}
    \includegraphics[width=0.49\textwidth, clip=True, trim=0.3cm 0.3cm 0.3cm 0.4cm]{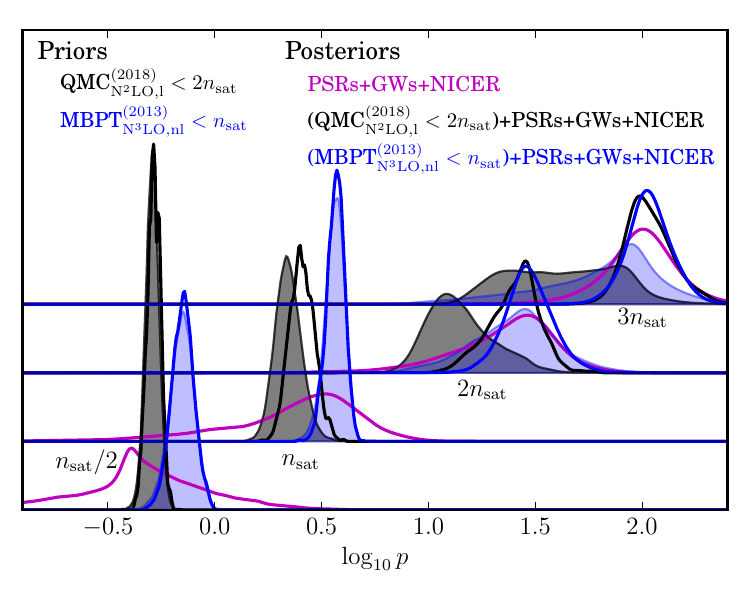}
    \caption{
        Prior (shaded) and posterior (lines) distributions for the pressure (in $\mathrm{MeV}/\mathrm{fm}^3$) at $0.5$, $1$, $2$, and $3\rhonuc$ when conditioning on data from only \PSRs~and \GWs~(left) or \PSRs, \GWs, and \NICER~observations (right).
        Each cluster of vertically-offset curves represents the distributions for the pressure at a specific density, with higher densities centered at higher pressures.
        We show both the prior conditioned on \QMCbasic~\EFT~calculations with local chiral interactions up to $2\rhonuc$ (\QMC; shaded grey region) and the prior conditioned on \MBPT~\EFT~calculations with nonlocal interactions up to $\rhonuc$ (\MBPT; shaded blue region), both with nuclear-theory agnostic extensions to higher densities.
        We also show posteriors obtained from a completely nuclear-theory agnostic analysis of the astrophysical data~\cite{Landry:2020} (red/magenta lines) as well as those obtained after initially conditioning on the \QMC~up to $2\rhonuc$ (black lines) and \MBPT~predictions up to \rhonuc~(blue lines).
        Our results quantify the extent to which the \EFT~calculations agree with the completely agnostic posterior, and therefore the astrophysical data.
    }
\label{fig:probability vs pressure}
\end{figure*}

Our study is enabled by a nonparametric representation of the \EOS~within a Bayesian inference scheme~\cite{Landry:2019, Essick:2019, Landry:2020}.
While we are not the first to incorporate \EFT~constraints in an analysis of astrophysical data, previous analyses employed \textit{ad hoc} parametric representations of the \EOS~at high densities, e.g., by using polytropic models~\cite{Hebeler:2010jx,Tews:2012fj,Hebeler:2013nza, Annala:2017llu, Most:2018hfd} or parametrizations of the speed of sound~\cite{Tews:2018chv, Greif:2018njt, Capano:2019eae, Raaijmakers:2019dks}.
These specific parametrization choices may introduce modeling systematics~\citep{Greif:2018njt} by restricting the allowed \EOSs~to a single functional family that may or may not match the true \EOS.
Furthermore, it can be difficult to assess the impact of prior beliefs for some parametrizations, complicating the interpretation of constraints \textit{a posteriori}.
Nonparametric \EOS~inference avoids both these issues by assigning nonzero prior probability to any causal and thermodynamically stable \EOS~according to transparent priors on correlations between the sound speed at different densities.
Our analysis, then, utilizes the most model freedom in the \EOS~to date and additionally incorporates the full uncertainty from \EFT~predictions, including uncertainty in the maximum density up to which \EFT~is valid.
Fig.~\ref{fig:probability vs pressure} demonstrates the intuition behind our results by comparing \EFT~predictions (shaded regions) with posterior distributions conditioned on astrophysical data.

In Sec.~\ref{sec:methodology} we provide a brief review of our \EFT~calculations along with a description of our nonparametric \EOS~inference methodology and incorporation of astrophysical data.
Sec.~\ref{sec:results} presents our main results, where we provide constraints on the maximum density and pressure up to which \EFT~calculations reproduce the astrophysical data.
We find that \EOS~models that include \EFT~constraints up to any density below $2\rhonuc$ are favored over more agnostic priors.
Astrophysical data slightly favor models that trust \EFT~up to $\sim2\rhonuc$ over those that trust \EFT~up to only $\rhonuc$ if only massive \PSRs~and \GW~observations are considered, but disfavor the former by a factor of $\lesssim2$ if the recent \NICER~measurements are also taken into account.
We additionally obtain updated constraints on the \EOS~by marginalizing over the uncertainty in the \EFT~breakdown density.
Sec.~\ref{sec:discussion} explores possible implications and caveats.

%--------------------------
\section{Methodology}
\label{sec:methodology}

We briefly review the \EFT~calculations we use in this work in Sec.~\ref{sec:EFT}, including both the \QMCbasic~calculations with local chiral interactions of Refs.~\cite{Gezerlis:2013ipa,Lynn:2015jua, Tews:2018kmu} and the \MBPTbasic~calculations of Ref.~\cite{Tews:2012fj} using nonlocal \EFT~interactions.
We then summarize the nonparametric \EOS~inference introduced in Refs.~\citep{Landry:2019, Essick:2019, Landry:2020} in Sec.~\ref{sec:GPR EOS}.

%------------
\subsection{Chiral effective field theory}
\label{sec:EFT}

Chiral EFT~\cite{Epelbaum:2008ga,Machleidt:2011zz} provides a systematic way to organize microscopic interactions between nucleons.
It starts from the most general Lagrangian that is consistent with the symmetries of \QCD, given in terms of pion and nucleon degrees of freedom.
This general set of interactions is then expanded in powers of a soft scale, either the external momentum or the pion mass, over a hard scale, the so-called breakdown scale $\Lambda_b$, that describes when additional degrees of freedom become important.
Individual contributions are arranged by their importance according to a so-called power counting scheme.
The result is a systematic order-by-order scheme for nuclear interactions, including both two- and many-body forces, that is typically truncated at third (next-to-next-to-leading order, N$^2$LO) or fourth order (next-to-next-to-next-to-leading order, N$^3$LO).
All the unknown coefficients, called low-energy couplings (LECs), are then fit to experimental data, i.e., nucleon-nucleon scattering data, binding energies or radii of light to medium-mass atomic nuclei, beta-decay matrix elements, or few-nucleon scattering observables.
Nuclear interactions obtained from \EFT~can then be employed in state-of-the-art many-body methods to solve the Schr\"odinger equation for nuclear systems.
For nucleonic matter, such methods include, for example, \QMCbasic~methods~\cite{Carlson:2015}, \MBPTbasic~\cite{Hebeler:2010jx, Wellenhofer:2014hya, Holt:2016pjb, Drischler:2017wtt}, the coupled-cluster method~\cite{Hagen:2013yba}, and the self-consistent Green's function method~\cite{Carbone:2013rca, Carbone:2013eqa}.

Although the truncation of the chiral expansion introduces theoretical errors~\cite{Epelbaum:2014efa, Furnstahl2015, Melendez:2017phj}, the order-by-order scheme bounds these errors.
By carefully estimating the expected error introduced by the truncation of the interactions from order-by-order calculations~\cite{Epelbaum:2014efa}, \EFT~provides not only a prediction for the \EOS~but also reliable theoretical uncertainty estimates.
This is the main advantage of \EFT~over more phenomenological treatments of the nuclear microphysics.
As capturing such theoretical uncertainties is one of the main motivations behind the nonparametric \EOS~inference scheme first developed in Refs.~\citep{Landry:2019, Essick:2019, Landry:2020}, \EFT~provides a natural framework to combine information from nuclear theory with a nonparametric representation of the \EOS~when analyzing astrophysical data.

In this work, we specifically use the \EFT~predictions for hadronic matter in beta equilibrium up to $2\rhonuc$ from Ref.~\cite{Tews:2018kmu}, which we refer to as \QMC.
These results were obtained by combining NL$^2$O chiral interactions in their local formulation from Refs.~\cite{Gezerlis:2013ipa,Lynn:2015jua} with \QMCbasic~methods, in particular with the auxiliary-field diffusion Monte Carlo (AFDMC) method~\cite{Schmidt:1999,Carlson:2015}.
The combination of \EFT~interactions with precise \QMCbasic~calculations produces binding energies and radii in light- to medium-mass nuclei in excellent agreement with experiments~\cite{Lonardoni:2017hgs, Lynn:2019rdt}.
These interactions have also been used to study neutron matter~\cite{Lynn:2015jua,Tews:2018kmu}, symmetric nuclear matter~\cite{Lonardoni:2019ypg}, and neutron-star matter~\cite{Tews:2018kmu} with great success.

However, we stress that there are other approaches to the \NS~\EOS~that start from microscopic calculations using different \EFT~Hamiltonians and many-body methods.
In particular, when implementing \EFT~interactions in many-body methods, regularization schemes need to be employed that remove high-momentum divergences.
These regularization schemes depend on the functional form of the regulator and a so-called cutoff parameter that describes which momentum scales are removed.
The \QMC~results employ local regulator functions at a cutoff scale of $R_0=1.0$ fm, which corresponds to a momentum-space cutoff of approximately $500$ MeV.
To explore the dependence of our results on the particulars of the microscopic approach, we also consider the neutron-matter \MBPTbasic~calculation of Ref.~\cite{Tews:2012fj}.
This calculation uses chiral interactions in their nonlocal formulation as input: two-nucleon forces at N$^3$LO from Refs.~\cite{Entem:2003ft, Epelbaum:2004fk}, as well as N$^3$LO many-body forces~\cite{Bernard:2007sp, Bernard:2011zr, Epelbaum:2006eu}.
These interactions employ nonlocal regulator functions with cutoffs in the range of 400--500 MeV.
We refer to this calculation as \MBPT.
Hence, the two approaches differ in both their \EFT~interactions and in the method used to solve the many-body problem, but we expect differences to be due primarily to the different nuclear interactions.

Importantly, while both \MBPT~and \QMC~predictions come with uncertainties, the particular \MBPTbasic~calculations of Ref.~\cite{Tews:2012fj} to date do not provide the same order-by-order error estimates as the \QMC~predictions.
Instead, \MBPT~estimates the error from the spread of results when using different nuclear Hamiltonians, cutoff scales, and three-nucleon LECs, while also accounting for the uncertainty due to the many-body method, but more recent \MBPTbasic~calculations also implement more advanced error estimation~\cite{Drischler:2017wtt,Drischler:2020hwi}.
%For the \QMC~calculation, the latter can be neglected.
Accordingly, the \QMC~error estimate includes the truncation uncertainty obtained from an order-by-order calculation, as well as uncertainties due to regulator artifacts.
In this sense, the uncertainty estimate of \QMC~is more systematic than for \MBPT, but at N$^3$LO the \MBPT~uncertainty estimate is nonetheless a reasonable approximation. 
Because theoretical uncertainty plays a key role in our analysis, we use the more systematic \QMC~results as our fiducial \EFT~predictions.

In addition to prior credible regions for the pressure at specific densities, calculations using \EFT~interactions determine the correlations between pressures at different densities by imposing the strong prior that the \EOS~is smooth within the regime where \EFT~is valid.
The \NS~\EOSs~predicted by \EFT, when one requires beta equilibrium and includes a consistent NS crust~\cite{Tews:2016ofv}, are well described by a parametrized model for the pressure as a function of the total energy density near \rhonuc:
\begin{multline}\label{eq:fit}
     p(\epsilon)\ [\mathrm{MeV}/\mathrm{fm}^3] = a_1 \left(\frac{\epsilon}{\epsilon_0}\right)^{\alpha} + b_1 \left(\frac{\epsilon}{\epsilon_0}\right)^{\beta} \\
         + c_1 \left(\frac{\epsilon}{\epsilon_0}\right)^{\gamma} e^{-\left(\epsilon/\epsilon_c\right)^2}
\end{multline}
with $\epsilon_0 = 150.227\ \mathrm{MeV}/\mathrm{fm}^3$.
We match each \EOS~realization to a low-density outer crust (\texttt{sly}~\citep{Douchin2001}) at densities below approximately 0.1$\rhonuc$, using the full \EFT~uncertainty estimates down to the fixed outer crust.
The \EFT~predictions already include an inner crust, and the assumption of a fixed outer crust at extremely low densities is not expected to affect our conclusions.
For both the \QMC~and \MBPT~calculations, we generate $\mathcal{O}(10^4)$ parameter sets that fall within the expected uncertainty band from each prediction.
There are nontrivial correlations between the parameters in Eq.~\eqref{eq:fit}, but \EFT~calculations typically produce pressures of $\mathcal{O}(10)\,\mathrm{MeV}/\mathrm{fm}^3$ near $2\rhonuc$.
As described in Sec.~\ref{sec:GPR EOS}, we condition our nonparametric representations of the \EOS~on the full prior process described by the \EFT~calculations, including all correlations between pressures at different densities, thereby capturing all the available theoretical information.

While \EFT~is a powerful tool, it cannot be applied at all densities probed in the interior of \NSs.
In particular, because it is an expansion in terms of momenta, \EFT~has a radius of convergence that is related to the breakdown scale.
As the density increases, nucleon momenta become comparable to $\Lambda_b$, and the expansion becomes less reliable as the details of the interactions at short distances become important.
This produces a rapid increase in the theoretical uncertainty associated with the \EOS~as the density increases.
Furthermore, many-body effects may also influence the radius of convergence in \NS~matter, and it is possible that phase transitions to matter containing degrees of freedom not included in \EFT~appear at densities between 1--2 $\rhonuc$.
Hence, the density (or pressure) at which \EFT~breaks down---and the nature of this breakdown---are unknown.
When excluding the possibility of a phase transition, the \QMC~calculation's order-by-order convergence suggests that the local chiral interactions are not predictive at densities above $2\rhonuc$~\cite{Tews:2018kmu}, and it is expected that the \QMC~calculations may break down somewhere between 1--2$\rhonuc$.
No such study was performed for the \MBPT~calculations, and we consequently employ the \MBPT~results only at densities $\lesssim \rhonuc$.
Our primary objective is to use astrophysical data to test \EFT~predictions between 1--2 $\rhonuc$ in search of such a breakdown.

%------------
\subsection{Nonparametric equation of state inference}
\label{sec:GPR EOS}

There is a well-established literature studying the ability of various parametrizations of the \NS~\EOS~to constrain our knowledge of nuclear interactions at high densities.
Techniques range in complexity from comparing pairs of candidate \EOSs~directly~\cite{LIGOScientific:2019eut} to modeling the \EOS~with a sound speed parametrization~\cite{Alford:2013aca, Tews:2018kmu, Greif:2018njt}, a piecewise polytrope~\cite{ReadLackey2009,Hebeler:2013nza,RaithelOzel2016, PhysRevD.91.043002} or a spectral parametrization~\cite{Lindblom2010,LindblomIndik2012,LindblomIndik2014,PhysRevD.98.063004, Wysocki:2020myz}.
However, all these approaches must concern themselves with modeling systematics as, by construction, they only represent a small family of possible \EOSs~which may or may not match the true \EOS~realized in nature.
Nonparametric inference, on the other hand, does not assume a specific functional family \textit{a priori} and assigns nonzero prior probability to any causal and thermodynamically stable \EOS, thereby guaranteeing that the inference will not suffer from the same type of modeling systematics inherent in parametrizations with a finite number of parameters.
We employ the nonparametric \EOS~representation based on Gaussian processes introduced in Refs.~\citep{Landry:2019, Essick:2019, Landry:2020}.
Those studies, and references therein, provide a pedagogical introduction to Gaussian processes and their application in nonparametric \EOS~inference.
We provide only a brief review of the most relevant features here.

By working in an auxiliary variable as a function of pressure
\begin{equation}
    \phi(p) = \log\left(\frac{c^2}{c_s^2} - 1\right) ,
\end{equation}
where $c_s^2 = dp/d\epsilon$ is the speed of sound and $c$ the speed of light, any realization of $\phi(p)$ will manifestly correspond to a causal and thermodynamically stable \EOS.
We condition mixture models of Gaussian processes for $\phi(p)$ on the same collection of 50 candidate \EOS~from the literature as Refs.~\citep{Essick:2019, Landry:2020}, directly marginalizing over separate processes conditioned on \EOSs~that contain only hadronic matter, that contain hadronic and hyperonic matter, and that contain hadronic and quark matter.
Each process constitutes a generative model for the \EOS, from which we draw individual realizations or \textit{synthetic} \EOSs.
As we are interested in exploring the full spectrum of possible \EOSs, we generate \textit{agnostic} priors that depend only weakly on the candidate \EOSs~upon which they were trained, often generating behavior not seen in any of the candidate \EOSs.
As such, we obtain similar results when analyzing the priors based on separate compositions and only report results marginalized over composition.

Crucially, this work extends Refs.~\citep{Essick:2019, Landry:2020} by additionally conditioning directly on \EFT~uncertainty estimates up to a maximum pressure, denoted $\pmax$, when creating the Gaussian processes.\footnote{Because our implementation parametrizes $\epsilon$ as a function of $p$ instead of the other way around, it is more natural to enforce maximum pressures than maximum densities when conditioning on theoretical \EFT~uncertainties. We report our findings for different $\pmax$ in terms of the approximate corresponding $n_\mathrm{max}=n(\pmax)$.}
We note that none of the 50 candidate \EOS~from the literature used to construct the agnostic priors in this work (or in previous nonparametric analyses) were computed within the \EFT~framework, and the inclusion of \EFT~priors constitutes novel theoretical information.
By treating the theoretical uncertainty from \EFT~on an equal footing with the collection of candidate \EOSs~from the literature, albeit with uncertainty estimates given by the \EFT~prediction rather than the large uncertainty from our agnostic hyperparameters (see Ref.~\citep{Essick:2019} for more details on hyperparameter optimization), we generate processes that automatically obey the \EFT~predictions at low densities and explore the full set of possible \EOSs~at high densities.
Although our nonparametric processes still formally support any causal and thermodynamically stable \EOS, this process results in a strong preference \textit{a priori} for synthetic \EOSs~that respect \EFT~predictions up to a maximum pressure scale.

To put this another way, we assign large model uncertainties to individual \EOS~predictions from the literature while allowing for rapid changes in the sound speed as a function of pressure, and simultaneously condition on the tight theoretical uncertainty from \EFT~at low densities.
The tight uncertainties from \EFT~restrict the conditioned Gaussian process to follow \EFT~predictions at low densities, while the broad modeling uncertainties assigned to the rest of the \EOS~proposals from the literature allow the process to explore all possible \EOS~phenomenology at high densities.
Additionally, by incorporating the full prior process from \EFT, we encode the full set of correlations expected from \EFT~within our Gaussian processes, which in turn produce synthetic \EOSs~that faithfully reproduce all aspects of \EFT~predictions at low densities, including the requirement that the \EOS~be smooth.

Refs.~\cite{Drischler:2020hwi, Drischler:2020yad} study similar ways to represent theoretical uncertainty from \EFT~with Gaussian processes.
In particular, they show how Gaussian processes can accurately represent theoretical uncertainty in neutron matter at densities below $2\rhonuc$, exploring the impact of the term-by-term convergence expected from \EFT~and correlated uncertainties at different densities.
Here, we focus on representing the uncertainty in the energy density as a function of pressure in beta-stable matter, capturing the correlations between multiple densities within the \QMC~and \MBPT~calculations we study, but it would be interesting to combine both approaches in future work.

Although an examination of the order-by-order convergence in the \EFT~calculation themselves provides information about $\pmax$, the uncertainty associated with the breakdown scale and the broad range of momenta involved in nuclear interactions in dense matter introduce intrinsic uncertainty in $\pmax$.
For this reason, we construct a set of processes conditioned on \EFT~predictions up to various $\pmax$, which in turn allows us to construct a posterior over both the \EOS~and $\pmax$ simultaneously.
All our processes retain broad uncertainties for the pressure at high densities, characteristic of the extreme model freedom in our agnostic nonparametric inference.
In this way, we can precisely quantify up to which $\pmax$ \EFT~predictions agree with astrophysical data without worrying about possible modeling systematics from \textit{ad hoc} parametric extensions to high densities.

Along those lines, our nonparametric priors do not impose strong correlations between the pressures at different densities above \pmax.
We therefore show how the astrophysical data itself correlates our knowledge of the pressure at different densities \textit{a posteriori}, and how decreasing our uncertainty at low densities by conditioning on \EFT~predictions up to a specific $\pmax$ can improve our knowledge of the \EOS~at pressures significantly above $\pmax$.

%-----------
\subsection{Incorporation of Astrophysical Observations}
\label{sec:astro likelihoods}

We incorporate astrophysical data following the prescriptions in Ref.~\cite{Landry:2020}, which show how to properly incorporate the full observational likelihoods from \GW, \NICER, and \PSR~data sets in a hierarchical Bayesian inference with nonparametric \EOS~priors.
We use the same \GW, \NICER, and \PSR~data as Ref.~\cite{Landry:2020} and assume the same fixed population models.
We explicitly incorporate the Occam factors from population models that truncate at $M_\mathrm{max}$ for \PSR~observations.
That is, we use Ref.~\cite{Landry:2020}'s Eqn. 5 with publicly available posterior samples for GW170817~\cite{TheLIGOScientific:2017qsa}, Eqn. 8 for \NICER~observations of J0030+0451~\citep{Miller:2019cac}, and Eqns. 10 and 11 for \PSR~mass measurements~\cite{Demorest:2010bx,Antoniadis:2013pzd,Cromartie:2019kug}.
We refer the reader to the discussion in Ref.~\cite{Landry:2020} for more detail.

%--------------------------
\section{Results}
\label{sec:results}

By exploring the impact of \EFT~predictions within a nonparametric \EOS~inference framework, we detail the ability of astrophysical observations to directly test and, in principle, validate microscopic calculations of dense neutron-rich matter.
Sec.~\ref{sec:max density} explores the ability of astrophysical data to constrain $\pmax$, while Sec.~\ref{sec:constraints} provides updated constraints on the \EOS~obtained by marginalizing over the uncertainty in $\pmax$.
Throughout, we consider the following astrophysical observations: (i) the most massive known \PSRs~\cite{Demorest:2010bx,Antoniadis:2013pzd,Cromartie:2019kug}, (ii) \GW~data from GW170817~\cite{TheLIGOScientific:2017qsa}, and (iii) \NICER~X-ray timing observations of J0030+0451~\citep{Miller:2019cac}.
We do not consider GW190425 as it is too quiet to inform the \EOS~\cite{Abbott:2020uma, Landry:2020}.

%------------

\subsection{Evidence for $\boldsymbol{\chi}$EFT and its breakdown scale}
\label{sec:max density}

\begin{figure*}
    \includegraphics[width=0.85\textwidth, clip=True, trim=0.7cm 0.4cm 0.7cm 0.3cm]{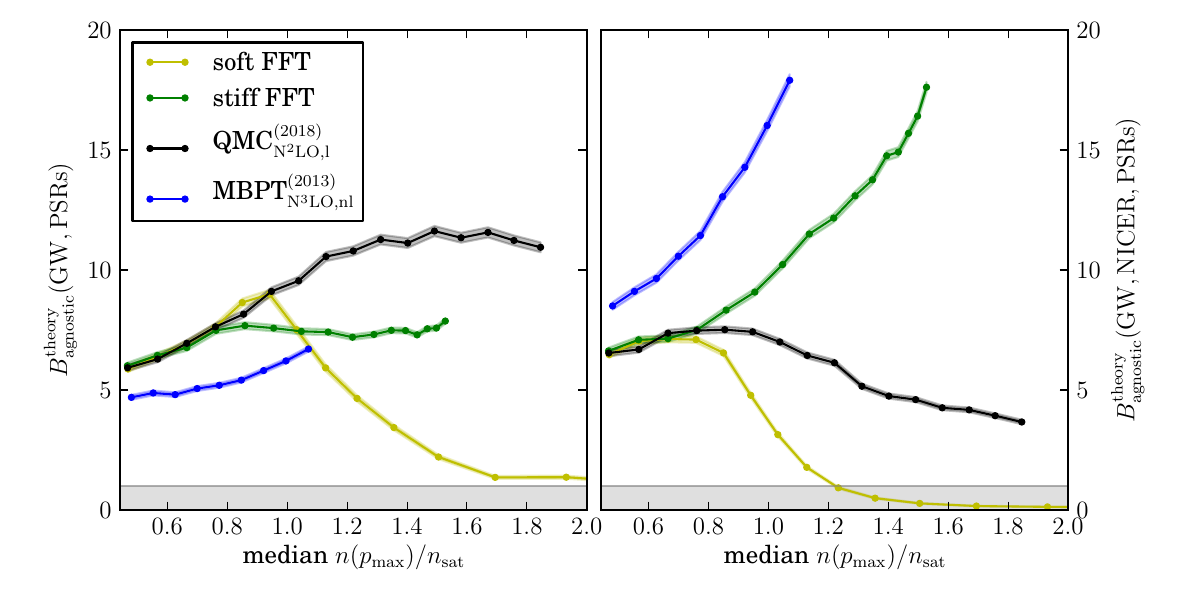}
    \caption{
        Bayes factors for astrophysical data conditioned on theoretical predictions up to different $\pmax$ compared to a completely agnostic analysis~\cite{Landry:2020} with only massive \PSRs~and \GW~data (left) and massive \PSRs, \GWs, and \NICER~observations (right).
        We plot the evidence ratios against the median density from the prior theoretical uncertainty at each $\pmax$.
        Both panels show the results assuming \QMC~calculations (black), \MBPT~calculations (blue), and artificial theories that are softer (yellow) or stiffer (green) than the \QMC~results at densities above $\rhonuc$.
        Shaded regions denote 1-$\sigma$ Monte-Carlo sampling uncertainties, and we highlight $\Bayes<1$ as the region where the prior with no information from nuclear theory is preferred over priors conditioned on nuclear theory results.
    }
    \label{fig:density vs evidence}
\end{figure*}

While we quantify the ability of astronomical data to test \EFT~predictions in some detail below, it is worthwhile to first obtain some intuition for these results.
Specifically, Fig.~\ref{fig:probability vs pressure} shows several prior (shaded regions) and posterior (thick lines) distributions for the pressure at a few densities, given different prior assumptions and astrophysical data sets (see Figs.~\ref{fig:corner GW} and~\ref{fig:corner NICER} in the Appendix for correlations between pressures).
Importantly, we show the results from a completely nuclear-theory agnostic analysis~\cite{Landry:2020} that was not conditioned on \EFT~in any way alongside the various \EFT~predictions.
A few trends are readily apparent.
The completely agnostic analysis retains relatively large uncertainty for all densities $\lesssim 2\rhonuc$ compared to \EFT~predictions.
While the agnostic analysis is consistent with \EFT~predictions, the statistical uncertainties are currently still too broad to confidently claim that the \EOS~has been measured precisely enough to validate \EFT.
However, the extent of the overlap between the \EFT~predictions and the agnostic constraints provides the simplest direct test of the \EFT~prediction.
To phrase this differently, \EFT~predictions agree to a remarkable degree with the most likely posterior predictions from agnostic nonparametric analyses of astrophysical observations that did not know about \EFT~results \textit{a priori}.

Fig.~\ref{fig:probability vs pressure} breaks down the constraints based on two sets of astrophysical data: massive \PSRs~and \GW~data alone vs. massive \PSRs, \GW, and \NICER~observations.
With only \PSRs~and \GW~data, we see that the \QMC~predictions for the pressure consistently fall near the maxima \textit{a posteriori} from the agnostic analysis, indicating that the data prefer \EOSs~that resemble the \QMC~result at low densities.
This trend is also true of the \MBPT~predictions, although they typically fall at slightly higher pressures than those most preferred by the data.
When we also include \NICER~observations, the peaks for the agnostic analysis shift to higher pressures at nearly all densities, reducing the extent to which the \QMC~results agree with the maxima \textit{a posteriori} (particularly above $\rhonuc$) and increasing the agreement between the data and \MBPT~predictions.
We stress that the completely agnostic distributions are not inconsistent with either of the two \EFT~predictions, but the relative locations of these distributions' peaks are suggestive.

This intuition is quantified in Fig.~\ref{fig:density vs evidence}, where we show the Bayes factor between the \EFT-informed models and the completely agnostic analysis.
Again, we break down our results based on whether or not we include \NICER's observations.
We construct a series of nested models for each \EFT~calculation which faithfully reproduce the theoretical predictions up to higher and higher $\pmax$ (see Fig.~\ref{fig:pressure-density}), and then compare them to the astrophysical data.
As with previous nonparametric analyses, we evaluate marginal likelihoods via direct Monte-Carlo integration over optimized Gaussian kernel density estimates of the likelihood, assuming uniform mass priors.
Following the procedures detailed in Ref.~\cite{Landry:2020}, we assume a fixed population of compact objects and therefore ignore selection effects, an acceptable approximation due to the small number of observed systems~\cite{Wysocki:2020myz}.
Fig.~\ref{fig:density vs evidence} shows the resulting Bayes factor between the models conditioned on \EFT~and the completely agnostic analysis, \Bayes.
A Bayes factor larger than unity implies that a model conditioned on \EFT~is preferred over the completely agnostic model.
Because there is uncertainty in the exact mapping between pressure and density,\footnote{For a given $\pmax$, the corresponding densities can vary between 6-17\% around the corresponding median density.} Fig.~\ref{fig:density vs evidence} reports the median density from each theory's prior uncertainty at each $\pmax$.

Fig.~\ref{fig:density vs evidence} shows that priors conditioned on either the \QMC~or \MBPT~results are favored over the agnostic prior regardless of the precise $\pmax$ up to which we trust \EFT.
The detailed behavior, however, depends on the astrophysical data we include.
When we consider only \PSRs~and \GWs, \EFT~predictions agree very well with the data up to as much as $2\rhonuc$.
This can be understood as an Occam factor.
The \EFT~predictions, unlike the agnostic analysis, limit the prior volume to regions of high likelihood and therefore produce larger marginal likelihoods.
Both the \QMC~and \MBPT~results follow this trend, and \Bayes~ increases when we trust each theory up to higher densities.
However, the softer \QMC~predictions are favored over the \MBPT~results since the \QMC~predictions fall closer to the maxima \textit{a posteriori} of the agnostic analysis.
Unfortunately, we do not have \MBPT~predictions beyond $\sim \rhonuc$ and, hence, cannot compare both approaches up to $2\rhonuc$.
For the \QMC~results, \Bayes~levels off at $\sim1.3\rhonuc$.
This indicates that further reduction in the prior volume from extending the \EFT~predictions to higher densities does not change the evidence, suggesting that the \PSR~and \GW~likelihoods are relatively flat at this scale.
The agreement between \EFT~predictions and the completely agnostic nonparametric analysis is remarkable in and of itself, although we stress again that the preference does not constitute absolute proof that \EFT~calculations do not break down below $\sim 1.3\rhonuc$.

What's more, the monotonic increase in \Bayes~with increasing $\pmax$ was not guaranteed.
As a demonstration, we construct two \textit{fake field theories} (FFTs), one that artificially softens (soft FFT) and one that stiffens (stiff FFT) relative to the real \QMC~predictions; Appendix~\ref{sec:fake field theories} describes the FFTs in more detail.
Both the soft and stiff FFTs retain the correlations between pressures predicted by the \QMC~calculations and follow the calculation's mean sound speed up to \rhonuc; the soft and stiff FFT processes only deviate from \QMC~results above \rhonuc, where they are centered on smaller and larger sound speeds, respectively.
Fig.~\ref{fig:density vs evidence} shows that astrophysical data is able to distinguish between the FFTs and the \QMC~results, just as it is able to distinguish between the \QMC~and \MBPT~predictions.
We also find that the astrophysical data correctly identifies the approximate density at which the FFTs begin to deviate from the \QMC~prediction, although the relative preference is not overwhelming.
For example, trusting the soft FFT up to $2\rhonuc$ is disfavored by a factor of \result{$\sim 6$} compared to trusting the same theory up to only \rhonuc, indicating that the former is hardly preferred over the completely agnostic model.
This demonstrates that astrophysical observations can, in principle, distinguish between different theories for nuclear interactions with high confidence, but smaller observational uncertainties are required before the statistical evidence will become overwhelming.
Furthermore, it is possible to recover the $\pmax$ up to which a theoretical prediction reproduces the observed data, which manifests as a local maximum in Fig.~\ref{fig:density vs evidence}.
The more pronounced this maximum is, the stronger the evidence for a breakdown of the theory.

While \EFT~predictions are still preferred over the agnostic analysis when we additionally include \NICER~observations, we see very different behavior for \Bayes.
\NICER~observations favor somewhat stiffer \EOSs~than are inferred from the \PSR~and \GW~data alone.
This tends to favor the stiffer \MBPT~predictions over the softer \QMC~predictions by at most a factor of \result{$\sim 2$}, even when we stop trusting either theory at a density below $\rhonuc$.
In fact, the preference of \NICER~data for stiff \EOSs~drives \Bayes~for both the \MBPT~results and the (completely artificial) stiff FFT to a factor of nearly \result{20} over the agnostic analysis.
Conversely, we find a local maximum for the \QMC~result, suggesting that the corresponding \EFT~predictions may begin to break down at, or just above, \rhonuc~because they are softer than the \textit{maxima a posteriori} from the completely agnostic analysis.
We emphasize, however, that the data's preference for the \QMC~results only falls by a factor of $\lesssim 2$ when we trust those predictions up to $2\rhonuc$.
As a more extreme example, the soft FFT is even more strongly disfavored, becoming less likely than the completely agnostic model when we trust it beyond $\sim1.2\rhonuc$.

Although the combination of massive \PSRs, \GWs, and \NICER~observations currently suggests that the true \EOS~is stiffer than the \QMC~prediction, and more like the \MBPT~result, this is far from definitive.
We obtain \Bayes~of no more than $\mathcal{O}(10)$, which is not large enough to be conclusive~\cite{Raftery:1995}.
The Bayes factors between the different \EFT~predictions are typically even smaller.
In fact, comparing the left and right panels of Fig.~\ref{fig:density vs evidence} makes it readily apparent that another observation could significantly alter the evidence.
Therefore, while intriguing, the current astrophysical data can not definitively suggest that \QMC~predictions break down at \textit{any} density below $\sim 2\rhonuc$.
For that matter, the current data cannot confidently show that we should trust \EFT~results even below \rhonuc, although the models conditioned on \EFT~up to $\sim\rhonuc$ are consistently favored over the agnostic prior by a factor of \result{$\gtrsim 7$}.

Nonetheless, as statistical uncertainties shrink with further astrophysical observations, we can expect strong tests of \EFT~in the density range between 1--2$\rhonuc$.
Indeed, this highlights the power of combining a full accounting of theoretical uncertainties with the extreme model freedom enabled by nonparametric analyses when testing nuclear interactions.

%------------

\subsection{Updated constraints on the high-density \EOS~ and \NS~structure}
\label{sec:constraints}

\begin{figure*}
    \includegraphics[width=1.0\textwidth, clip=True, trim=0.5cm 0.0cm 0.5cm 0.1cm]{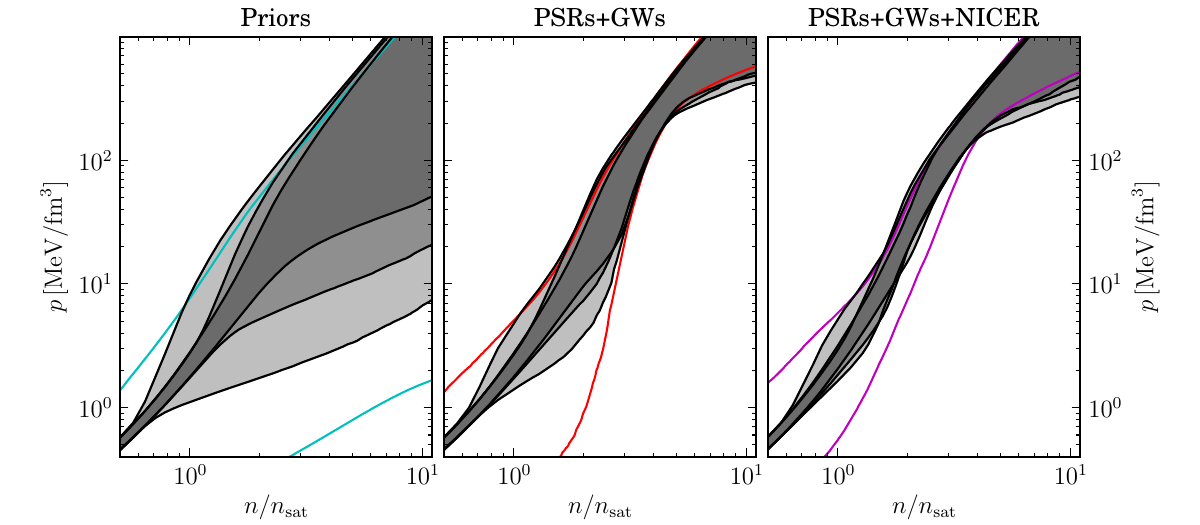}
    \includegraphics[width=1.0\textwidth, clip=True, trim=0.5cm 0.0cm 0.5cm 0.6cm]{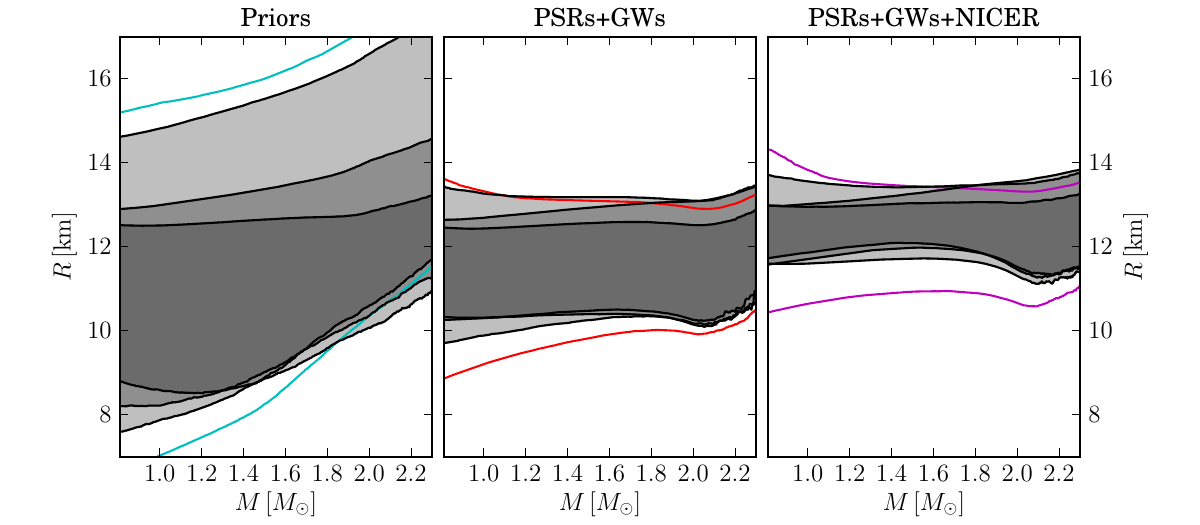}
    \caption{
        90\% symmetric credible regions for the (top) pressure at each density and (bottom) radius at each mass.
        We show (left) prior processes conditioned on only \QMC~predictions up to approximately (from lighter shading to darker shading) $0.5\times$, $1\times$, and $2\rhonuc$, (middle) posteriors conditioned on massive \PSRs~and \GW~data, as well as (right) posteriors conditioned on massive \PSRs, \GWs, and \NICER~observations.
        In all panels, analogous results obtained with a completely agnostic analysis not conditioned on \EFT~\cite{Landry:2020} are shown in as colored lines.
    }
    \label{fig:pressure-density}
\end{figure*}

In addition to demonstrating the ability of astrophysical observations to constrain the density and pressure scales up to which we should trust theoretical predictions, we marginalize over the uncertainty in $\pmax$ to obtain updated constraints on \NS~properties and the dense-matter \EOS.
Since Fig.~\ref{fig:probability vs pressure} shows that the posteriors obtained from \QMC~and \MBPT~calculations are quantitatively similar, in what follows we focus on the results conditioned on the \QMC~predictions because of their more systematic order-by-order error estimation.

Existing analyses based on \EFT~predictions report posterior credible regions conditioned on (at most) a few prior choices of $\pmax$.
This immediately raises the question of which prior, and therefore posterior, one should trust.
Using our sets of nested models, we construct a posterior distribution over both the \EOS~and $\pmax$ simultaneously.
By marginalizing over the posteriors conditioned on separate $\pmax$, sometimes called \textit{model averaging}, we allow the data itself to not only tell us the relative likelihood of each $\pmax$ but automatically incorporate the $\pmax$ uncertainty into the posterior for any other quantity we wish.
Similar approaches have been proposed to mitigate theoretical uncertainty in, e.g., \GW~waveforms~\cite{PhysRevD.101.064037}.
To wit,
\begin{align}
    P(X|\mathrm{data}) & = \int d\pmax\, P(X,\pmax|\mathrm{data}) \nonumber \\
           & = \int d\pmax\, P(X|\mathrm{data},\pmax) P(\pmax|\mathrm{data}) \nonumber \\
           & \propto \int d\pmax\, \left[ P(X|\mathrm{data},\pmax) \right. \nonumber \\
           & \quad\quad\quad\quad\quad\quad \left. \times P(\mathrm{data}|\pmax)P(\pmax) \right] ,
\end{align}
and we see that if the data strongly prefer a single $\pmax$, implying that only a single $P(\mathrm{data}|\pmax)$ is large, the model-averaged posterior for a given observable $X$ collapses to the posterior conditioned on the preferred $\pmax$.

\begin{table*}
    \centering
    \begin{tabular}{c||c|c||c|c||c|c||c|c}
        \hline
         \multirow{2}{*}{Quantity} &  \multicolumn{2}{c||}{\QMC$\lesssim \rhonuc$} & \multicolumn{2}{c||}{\QMC$\lesssim 2\rhonuc$} & \multicolumn{2}{c||}{Marginalized \QMC} & \multicolumn{2}{c}{Completely Agnostic} \\
         \cline{2-9}
                                   & \PSRs+\GWs & +\NICER & \PSRs+\GWs & +\NICER & \PSRs+\GWs & +\NICER & \PSRs+\GWs & +\NICER \\
         \hline
         \hline
         $M_\mathrm{max}\ [M_\odot]$ & \PSRGWMmaxOne & \PSRGWNICERMmaxOne & \PSRGWMmaxTwo & \PSRGWNICERMmaxTwo & \PSRGWMmaxMrg & \PSRGWNICERMmaxMrg & \PSRGWMmaxAgn & \PSRGWNICERMmaxAgn \\
         $R_{1.4}\ [\mathrm{km}]$    & \PSRGWROne    & \PSRGWNICERROne    & \PSRGWRTwo    & \PSRGWNICERRTwo    & \PSRGWRMrg    & \PSRGWNICERRMrg & \PSRGWRAgn & \PSRGWNICERRAgn\\
         $\Lambda_{1.4}$             & \PSRGWLOne    & \PSRGWNICERLOne    & \PSRGWLTwo    & \PSRGWNICERLTwo    & \PSRGWLMrg    & \PSRGWNICERLMrg & \PSRGWLAgn & \PSRGWNICERLAgn \\
         \hline
         $p(\rhonuc)\ [\mathrm{MeV}/\mathrm{fm}^3]$         & \PSRGWPoneOne & \PSRGWNICERPoneOne & \PSRGWPoneTwo & \PSRGWNICERPoneTwo & \PSRGWPoneMrg & \PSRGWNICERPoneMrg & \PSRGWPoneAgn & \PSRGWNICERPoneAgn \\
         $p(2\rhonuc)\ [10\, \mathrm{MeV}/\mathrm{fm}^3]$   & \PSRGWPtwoOne & \PSRGWNICERPtwoOne & \PSRGWPtwoTwo & \PSRGWNICERPtwoTwo & \PSRGWPtwoMrg & \PSRGWNICERPtwoMrg & \PSRGWPtwoAgn & \PSRGWNICERPtwoAgn \\
         $p(3\rhonuc)\ [10^2\, \mathrm{MeV}/\mathrm{fm}^3]$ & \PSRGWPthreeOne & \PSRGWNICERPthreeOne & \PSRGWPthreeTwo & \PSRGWNICERPthreeTwo & \PSRGWPthreeMrg & \PSRGWNICERPthreeMrg & \PSRGWPthreeAgn & \PSRGWNICERPthreeAgn \\
         $p(4\rhonuc)\ [10^2\, \mathrm{MeV}/\mathrm{fm}^3]$ & \PSRGWPfourOne & \PSRGWNICERPfourOne & \PSRGWPfourTwo & \PSRGWNICERPfourTwo & \PSRGWPfourMrg & \PSRGWNICERPfourMrg & \PSRGWPfourAgn & \PSRGWNICERPfourAgn \\
         $p(6\rhonuc)\ [10^2\, \mathrm{MeV}/\mathrm{fm}^3]$ & \PSRGWPsixOne & \PSRGWNICERPsixOne & \PSRGWPsixTwo & \PSRGWNICERPsixTwo & \PSRGWPsixMrg & \PSRGWNICERPsixMrg & \PSRGWPsixAgn & \PSRGWNICERPsixAgn \\
         \hline
    \end{tabular}
    \caption{
        Median values and 90\% highest-probability-density credible regions for a few macroscopic observables and pressures at reference densities for posteriors conditioned on massive \PSRs~and \GWs~as well as \PSRs, \GWs, and \NICER~observations.
        We quote results for processes conditioned on \QMC~predictions up to $\sim\rhonuc$, $\sim2\rhonuc$, and marginalized over $\pmax$ with an approximately uniform prior over $n_\mathrm{max}$ between $\rhonuc/2$ and $2\rhonuc$.
        We also reproduce the results from a completely agnostic analysis not conditioned on \EFT~\cite{Landry:2020}.
    }
    \label{tab:credible regions}
\end{table*}

Table~\ref{tab:credible regions} presents median values and 90\% highest-probability-density credible regions for canonical macroscopic observables and for pressures at a few reference densities after conditioning on massive \PSRs, \GWs, and \NICER~observations, as well as \QMC~predictions at low densities.
In addition to quoting credible regions conditioned on individual choices of $\pmax$, we also present constraints marginalized over the uncertainty in $\pmax$.
As $P(\mathrm{data}|\pmax)$ varies by at most a factor of a few (see Fig.~\ref{fig:density vs evidence}), the model-averaged result may be influenced by our choice of $P(\pmax)$.
We impose a prior that is approximately flat in $n(\pmax)$ between $\rhonuc/2$ and $2\rhonuc$ as determined by the \QMC~predictions.
However, because the actual posteriors obtained with separate $\pmax$ are not radically different (see Fig.~\ref{fig:pressure-density}), this ambiguity may be moot, as the model averaging will average nearly identical posteriors.

We note that the completely agnostic constraints shown in Figs.~\ref{fig:probability vs pressure} and \ref{fig:pressure-density} are relatively weak at low densities because the \NS~properties measured in astrophysical systems are not very sensitive to the \EOS~at densities below $\rhonuc$.
The agnostic constraints are stronger at densities around 2--3 $\rhonuc$, as the pressures in this regime have a more pronounced impact on \NS~structure for stars with gravitational masses between 1--2 $M_\odot$~\cite{Weih_2019}.

However, it is also apparent that the posteriors conditioned on \EFT~results are narrower than those obtained from a completely agnostic analysis.
This is likely not exclusively due to the tighter \EFT-informed priors, although they clearly play a role, because the uncertainties at high densities are as broad as those from the agnostic analysis.
We also see differences \textit{a posteriori} between the different \EFT-informed priors, particularly at low densities.
The tighter constraints, then, are due to the integrated nature of the macroscopic observables, which are sensitive to the entire \EOS~up to some central density and pressure.
In other words, it is the likelihood that is responsible for correlating pressures at different densities, rather than the prior.
As such, improved knowledge of the low-density \EOS~translates into better knowledge at higher densities; in particular, more stringent prior constraints at low densities combine with the likelihood's correlations to produce tighter posterior bounds at higher densities.

We also note that the process conditioned on \QMC~up to 2\rhonuc~produces slightly higher pressures at \rhonuc~\textit{a posteriori} than if we condition on \QMC~results up to only \rhonuc.
In fact, the 2\rhonuc~posterior seems to be centered at larger pressures than the \QMC~prior.
This is due to the rigidity of the \EFT~predictions when we condition on \EFT~up to large \pmax, which require smooth \EOSs, and the likelihood's strong preference for larger pressures at higher densities, driven primarily by the \NICER~observations.
In order to satisfy the likelihood at high densities, the data favor the stiffer part of the \EFT~prior range.
This also increases the pressure at low densities because \EFT~calculations predict strong correlations between pressures at different densities.
Taken to an extreme, this resembles the type of modeling systematics associated with parametrized functional families.
However, in the limit of many observations, our nonparametric analysis would eventually pick the correct \EOS~because our prior still supports every possible curve, unlike an actual parametrized analyses.
Nonetheless, we stress that strong prior assumptions about the functional form of the \EOS, like those present in the model conditioned on \EFT~up to 2$\rhonuc$, can introduce unexpected behavior \textit{a posteriori} because the prior correlates observables in a way that is difficult to disentangle.

One possible breakdown mechanism for \EFT~is a phase transition, and a possible signature of phase transitions, if they are strong enough, is disconnected hybrid-star branch in the mass-radius relation.
Although not all first-order phase transitions lead to more than one stable branch, the presence of multiple branches indicates a strong phase transition.
Given our prior processes conditioned on \EFT~results, the astrophysical data disfavor \EOSs~with multiple stable branches because the vast majority of synthetic \EOSs~with multiple stable branches generated \textit{a priori} do not support $2\,M_\odot$ stars.
These are therefore incompatible with the existence of massive \PSRs.
This finding is consistent with previous studies~\cite{Essick:2019, Landry:2020}, and we recover the \GWs' weak evidence in favor of multiple stable branches only when we condition on the \PSR~data \textit{a priori}.
The difference originates entirely in what is included in the prior, i.e., whether it includes the \PSR~observations or not.
Additionally, the preference for multiple stable branches is suppressed as we condition on \EFT~results up to higher densities.

%------------------------

\section{Discussion}
\label{sec:discussion}

By combining the predictions of \EFT~with nonparametric \EOS~inference, we show that astrophysical observations can directly constrain the density range over which nuclear theory calculations may begin to break down. We find that \EOS~priors incorporating \EFT~predictions at supranuclear densities are preferred over a completely nuclear-theory agnostic prior regardless of the maximum pressure up to which we trust \EFT.
In particular, we consider the evidence for a possible breakdown in two theoretical calculations, both of which are based on \EFT~interactions.
While \GWs~and massive \PSRs~favor the inclusion of \QMC~information up to nearly $2\rhonuc$, \NICER~observations suggest that the true \EOS~begins to stiffen compared to the \QMC~predictions at densities near or slightly above $\rhonuc$.
As a result, the inclusion of \NICER~data disfavors the model conditioned on \QMC~up to $2\rhonuc$ relative to the model conditioned only up to $\rhonuc$ by up to a factor of \result{$\sim 2$}.
In fact, \NICER's preference for stiffer \EOSs~better matches the predictions of the \MBPT~calculations with nonlocal interactions.
For this reason, it is interesting to consider the differences between the \QMC~and \MBPT~calculations and what this may tell us about fundamental nuclear interactions, assuming that the astrophysical observations are free of significant systematic errors.

The two calculations differ in the particulars of the many-body interactions they implement, as well as the computational many-body method.
It is reasonable to assume that at the cutoff range employed, the different \EFT~computations work reasonably well, and differences due to the many-body method can be neglected.
Thus, differences between the theoretical predictions must be due to the assumed interactions, i.e., the influence of the regularization scheme, cutoff range, the chiral order up to which the interactions are calculated, or the data to which the interactions are fit.

Our main results are based on the \QMC~calculations using local \EFT~interactions at a cutoff scale of $R_0=1.0$ fm.
It has been shown that local interactions lead to sizable regulator artifacts~\cite{Tews:2015ufa, Dyhdalo:2016ygz, Huth:2017wzw} that mainly affect long- and short-range parts of the three-body interactions.
In particular, local interactions lead to less repulsion from three-nucleon--pion-exchange interactions and the appearance of three-nucleon--contact regulator artifacts.
The \MBPT~results do not suffer from such large regulator artifacts as they use nonlocal chiral interactions.
This means that, while estimates for the uncertainty due to local contact regulator artifacts are included in the \QMC~uncertainty bands, those calculations may predict softer \EOSs~than \MBPT~because they include less repulsion from long-range three-nucleon interactions.
Additionally, both calculations only explore a limited range of cutoff values.
Future studies at larger cutoffs should reduce these artifacts, and allow for further investigations of our results' dependence on the regulator.

Furthermore, both calculations explore chiral interactions at different chiral orders.
The \QMC~calculation employs N$^2$LO chiral interactions, while the \MBPT~calculation uses interactions at N$^3$LO.
In principle, the difference should already be included in the order-by-order uncertainty estimates of the \QMC~calculation.
However, the N$^3$LO many-body forces have been found to be sizable~\cite{Tews:2012fj} and the data's preference for stiffer \EOSs~than \QMC~predicts could indicate that these forces do not follow the expected order-by-order convergence.
Future work similar to Refs.~\cite{Drischler:2017wtt,Drischler:2020hwi} is needed to determine the order-by-order behavior of the \NS~\EOS~up to N$^3$LO, including important many-body forces.
While it may be tempting to equate the stiffer predictions from \MBPT~with the higher-order many-body interactions included, it is important to remember that regulator artifacts could also affect the results.

The interactions employed here are all derived via Weinberg power counting~\cite{Weinberg1990, Weinberg1991}, which has several shortcomings~\cite{Tews:2020hgp}, among which are the dependence on the regularization scheme and scale.
Results using alternative power counting schemes, however, can easily be explored within the current framework, which allows us to investigate when any theoretical prediction begins to disagree with the astrophysical data.
Future work may compare a variety of \EFT-based calculations, i.e, using both local and nonlocal interactions at various chiral orders, exploring a similar density range, and using similar estimates for the theoretical uncertainty to reliably extract the breakdown density of \EFT~in \NS~matter.
While the current statistical precision from astrophysical data and theoretical uncertainties are not definitive at the moment, our framework ideally suited for this task.

Besides examining the evidence for \EFT~and its breakdown scale, we also provide updated constraints on the \EOS~that reflect both \EFT~predictions and the latest astronomical data.
We mitigate the sensitivity of our results to the specific choice of density up to which we trust \EFT~by marginalizing over $\pmax$.
We focus on the \QMC~predictions due to their more systematic error estimation and find comparable, but somewhat tighter, credible regions for NS observables as other studies~\cite{Miller:2019cac,Raaijmakers:2019dks,JiangTang2019,Landry:2020}.
For example, we constrain the radius and tidal deformability of a $1.4\,M_\odot$ \NS~to $R_{1.4}=\PSRGWNICERRMrg\,\mathrm{km}$ and $\Lambda_{1.4}=\PSRGWNICERLMrg$ with \PSR, \GW, and \NICER~observations, whereas previous nonparametric analyses that include the same observational data but did not condition on \EFT~\cite{Landry:2020} found \externalresult{$R_{1.4}=12.32^{+1.09}_{-1.47}$} and \externalresult{$\Lambda_{1.4}=451^{+241}_{-279}$}.
Similarly, we find $p(\rhonuc)=\PSRGWNICERPoneMrg\,\mathrm{MeV}/\mathrm{fm}^3$ and $p(2\rhonuc)=\PSRGWNICERPtwoMrgINLINE\,\mathrm{MeV}/\mathrm{fm}^3$, whereas Ref.~\cite{Landry:2020} found \externalresult{$p(\rhonuc)=2.68^{+2.37}_{-2.48}\,\mathrm{MeV}/\mathrm{fm}^3$} and \externalresult{$p(2\rhonuc)=23.8^{+16.6}_{-18.3}\,\mathrm{MeV}/\mathrm{fm}^3$}.
Conditioning on \EFT~predictions produces \EOSs~that are somewhat stiffer around $2\rhonuc$ compared to the \EOSs~preferred \textit{a posteriori} by the agnostic analysis, although the constraints agree within their uncertainties and the maxima \textit{a posteriori} are similar.
This is likely due to the strong theoretical prior introduced by \EFT~calculations at densities below $\rhonuc$, which rules out many of the synthetic \EOSs~allowed in the agnostic prior at lower densities.
Furthermore, our constraints on the very-high-density \EOS~are not improved by conditioning on \EFT~at low densities, as expected.
For example, we find $M_\mathrm{max}=\PSRGWNICERMmaxMrg\,M_\odot$, while Ref.~\cite{Landry:2020} reported \externalresult{$M_\mathrm{max}=2.22^{+0.30}_{-0.20}\,M_\odot$}.
Additionally, imposing the constraint that $M_\mathrm{max} \lesssim 2.2\,M_\odot$, as some authors have suggested based on the final fate of GW170817's remnant (see, e.g.,~\cite{Margalit&Metzger17, Rezzolla+18} although there is some debate about these constraints~\cite{Shibata:2019ctb}), has very little effect on our conclusions, both qualitatively and quantitatively.
Other studies have similarly observed insensitivity at low and intermediate densities to upper limits on $M_\mathrm{max}$~\cite{Capano:2019eae}.

Several previous studies of GW170817 incorporated \EOS~constraints from \EFT~\cite{Capano:2019eae,Tews2018,Tews:2019cap}, and it is instructive to compare their results to our constraints based on massive \PSR~and \GW~data.
Ref.~\cite{Capano:2019eae} obtained \externalresult{$R_{1.4} = 11.0^{+0.9}_{-0.6}\,$km} and \externalresult{$p(4n_{\rm sat}) = 161^{+58}_{-46}\, \mathrm{MeV}/\mathrm{fm}^3$} using \QMC$_{2018}$ information up to 2$n_{\rm sat}$ and a piecewise extension of the sound speed thereafter, with cuts based on $M_\mathrm{max}$ motivated by the existence of massive pulsars and the observations of GW170817's electromagnetic counterpart.
Our corresponding finding of $R_{1.4} = \PSRGWRTwo\,$km for \EFT~constraints below 2$n_{\rm sat}$ differs chiefly in its looser upper bound, which translates to a higher pressure of $p(4\rhonuc)=\PSRGWNICERPfourMrgMeV\,\mathrm{MeV}/\mathrm{fm}^3$.
The differences in our results are likely attributable to different choices of prior (Ref.~\cite{Capano:2019eae} favors smaller $R_{1.4}$ \textit{a priori}) and of high-density \EOS~extension.
Refs.~\cite{Tews2018,Tews:2019cap} adopted the same low-density model for the \EOS~as Ref.~\cite{Capano:2019eae} in their analysis of GW170817, but used the 90\% credible interval for $\tilde{\Lambda}$ in place of the full likelihood distribution over masses and tidal deformabilities, obtaining an upper bound of \externalresult{$R_{1.4}=12.6$ km}.

Like in the present work, Ref.~\cite{Raaijmakers:2019dks} performed an \EFT-informed joint analysis of data from massive \PSRs, GW170817, and \NICER's observation of \PSR~J0030+0451.
Using a different \EFT~prediction calculated with \MBPTbasic~up to 1.1$\rhonuc$ with a piecewise polytrope or parameterized sound speed extension at higher densities, they obtained posterior credible regions in the mass-radius plane centered on \externalresult{$\approx 12$--$12.5\,$km} at $1. 4\,M_{\odot}$, depending on the extension.
However, their posteriors are strongly influenced by their tight \EOS~prior, which excludes radii above \externalresult{13 km} at 90\% confidence in the case of the sound speed extension.
Despite our broader prior, we obtain a comparably tight posterior in the mass-radius plane.
Similarly, Ref.~\cite{Dietrich:2020lps} analyzed the same observations, approximating the \NICER~likelihood by its 2-$\sigma$ credible region and additionally incorporating information from lightcurve modeling of GW170817's kilonova counterpart, AT2017gfo~\cite{Abbott_2017} (see Ref.~\cite{10.1093/mnras/stz3457} for a discussion of kilonova modeling systematics).
They obtain \externalresult{$10.98^{+1.00}_{-0.69}$ km} for the canonical \NS~radius, which is smaller than other estimates that include \NICER~observations because their use of the 2-$\sigma$ credible region does not incorporate the full \NICER~likelihood.
These differences illustrate the constraints' level of sensitivity to the high-density representation of the \EOS, the way in which astrophysical likelihoods are modeled, and other prior choices.

Finally, it is worth considering what, exactly, we have learned from recent astrophysical observations.
It is readily apparent from Fig.~\ref{fig:pressure-density} that the combination of massive \PSR, \GW, and \NICER~observations can dramatically reduce the uncertainty in the \EOS.
However, much of the high-density information originates from the existence of massive \PSRs~(see Fig.~\ref{fig:fancy EFT process plots}), which have been known for nearly a decade.
Generally, we find our uncertainty in the macroscopic properties of canonical $1.4\,M_\odot$ stars is reduced by approximately a factor of two when we include \GW~and \NICER~data in addition to just \PSR~observations.
Unsurprisingly, \NICER~dominates the improvement in $R_{1.4}$, \GWs~have the most impact on $\Lambda_{1.4}$, but both contribute approximately equally to our final uncertainty in the \EOS.
Similarly, our knowledge of the pressure between 1--2\rhonuc~is improved, but typically to a smaller extent than the macroscopic observables.
This is likely due to the fact that multiple \EOS~can produce nearly the same $R$ and $\Lambda$ at a given mass.

\begin{figure}
    \includegraphics[width=1.0\columnwidth, clip=True, trim=0.3cm 0.2cm 0.3cm 0.3cm]{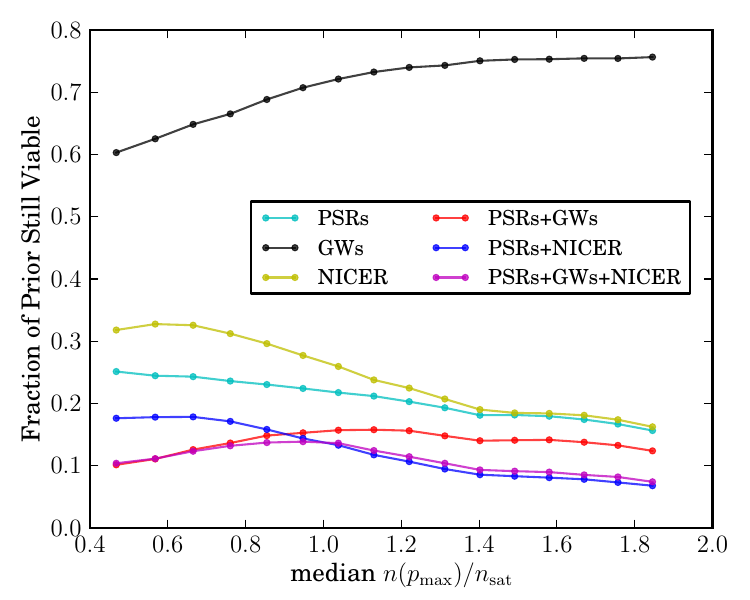}
    \caption{
        Fractions of the prior that are still viable after conditioning on astrophysical observations when trusting \QMC~up to different $\pmax$.
        More informative observations reduce our uncertainty relative to the prior, leaving only a fraction of the prior still viable (see Appendix~\ref{sec:information}).
    }
    \label{fig:information}
\end{figure}

To investigate the information gained by astrophysical observations, in Fig.~\ref{fig:information} we show the fraction of the prior that remains viable \textit{a posteriori} (based on the Shannon information~\cite{Shannon:1948:MTCa} in the posterior; see Appendix~\ref{sec:information}) when adding different combinations of data as a function of the maximum density up to which we condition on the \QMC~results.
Both \NICER~and massive \PSR~observations are very informative, although they individually prefer slightly different subsets of the prior.
\GW~observations alone, on the other hand, are not particularly informative, in agreement with previous studies that have shown that \GW~observations tend to reproduce the \QMC~predictions~\cite{Tews2018, Tews:2019cap}.
Therefore, adding \GW~information is redundant when we condition on \EFT~up to sufficiently high densities \textit{a priori}.
Conversely, \NICER~and massive \PSR~data become more informative as we trust \EFT~results up to higher densities, likely because they require relatively stiff \EOS~that differ from the \QMC~predictions.

Additional information can be obtained by combining different observations, and we find that the overall joint constraint follows the most-informative subset of astrophysical data.
We also note that the additional information gained by adding \NICER~observations to \GW~and \PSR~data is small when we trust \QMC~up to $\lesssim\rhonuc$, in agreement with Ref.~\cite{Raaijmakers:2019dks}.
Nonetheless, including \NICER~observations is at least as informative as adding \GW~observations to \PSR~data if we trust the \QMC~results up to higher densities.

Additionally, Ref.~\cite{Landry:2020} reports possible future constraints after approximately 1 year of \GW~detections at the current detectors' design sensitivities and 6 total \NICER~targets become available (expected by the mid 2020's).
Their projected uncertainty, which was not conditioned on any \EFT~results, is typically \externalresult{$\lesssim2$} times smaller than our current results for macroscopic observables of $1.4\,M_\odot$ stars and pressures near $2\rhonuc$.
We therefore can expect future astrophysical observations by themselves to be slightly more informative than our current observations conditioned on \EFT~predictions, with the notable exception of pressures at densities $\lesssim\rhonuc$.
At these low densities, \EFT~predictions are so tight that they will remain at least a factor of \externalresult{2} better than constraints obtained from astrophysical data alone in the foreseeable future.
Nonetheless, as statistical uncertainties continue to shrink, we can look forward to strong constraints on the possible breakdown of \EFT~calculations that can directly inform our understanding of when different underlying microphysics may become important.

While neither \EFT~nor more \GW~and \NICER~observations are likely to improve our knowledge of the \EOS~at extremely high densities, the path forward is clear for densities up to 3--4 $\rhonuc$.
By combining theoretical predictions from \EFT~with nonparametric \EOS~representations and marginalizing over $\pmax$, we will be able to exploit both observational constraints, i.e., \GW~and \NICER~observations that primarily improve our knowledge of the \EOS~between 1--3\rhonuc, while using theoretical information at lower densities, with the transition between theoretical predictions and direct astrophysical constraints dictated by the data rather than prior assumptions.
Such analyses will provide more stringent tests of \EFT~predictions and constrain the many-body forces between nucleons.
Indeed, as we have seen, \EFT-based many-body calculations can be used to construct a parameterization of the \EOS~up to 1-2\rhonuc, properly encoding correlations between the pressure at different densities suggested by theory within this interval.
We used a simple parameterization for convenience in this work, although others are perhaps better motivated or more flexible~\cite{Forbes:2019xaz}.
Additionally, Refs.~\cite{Drischler:2020hwi, Drischler:2020yad} represent theoretical uncertainty from \EFT~at densities below $2\rhonuc$ directly with Gaussian processes, similar to the low-density models based on our parametrization upon which we condition our prior processes.
Our framework can then directly constrain the maximum density up to which the theoretical parametrization matches the astrophysical data as well as the low-density theoretical parameters themselves.
Additionally, a focused effort in many-body theory could exploit the low-density \EOSs~preferred by astrophysical data to illuminate the nuclear forces that are challenging to constrain in the laboratory, similar to our current speculations with existing data.
This promises precise constraints on the \EOS, better understanding of \NS~in various astrophysical scenarios, and insight into the fundamental interactions governing nuclear matter.

%--------------------------
\acknowledgments

The authors thank Joseph Carlson, Christian Drischler, Amanda Farah, Maya Fishbach, as well as Katerina Chatziioannou and the LIGO-Virgo-KAGRA Extreme Matter working group for useful discussions while preparing this manuscript.
R.~E. and D.~E.~H. are supported at the University of Chicago by the Kavli Institute for Cosmological Physics through an endowment from the Kavli Foundation and its founder Fred Kavli.
D.~E.~H. is also supported by NSF grant PHY-1708081, and gratefully acknowledges the Marion and Stuart Rice Award.
P.~L. is supported by National Science Foundation award PHY-1836734 and by a gift from the Dan Black Family Foundation to the Gravitational-Wave Physics \& Astronomy Center.
I.~T. is supported by the U.S. Department of Energy, Office of Science, Office of Nuclear Physics, under contract No.~DE-AC52-06NA25396, by the NUCLEI SciDAC program, and by the LDRD program at LANL.
S.~R. was supported by the NSF grant PHY-1430152 to the JINA Center for the Evolution of the Elements and the Department of Energy grant DE-FG02-00ER41132.
The authors also gratefully acknowledge the computational resources provided by the LIGO Laboratory and supported by NSF grants PHY-0757058 and PHY-0823459.
Computational resources have also been provided by the Los Alamos National Laboratory Institutional Computing Program, which is supported by the U.S. Department of Energy National Nuclear Security Administration under Contract No.~89233218CNA000001, and by the National Energy Research Scientific Computing Center (NERSC), which is supported by the U.S. Department of Energy, Office of Science, under contract No.~DE-AC02-05CH11231.

%--------------------------

\bibliography{biblio}

%--------------------------
\appendix

\section{Additional Details of the Fake Field Theories and the Impact of Individual Observations}
\label{sec:fake field theories}

\begin{figure*}
    \includegraphics[width=1.0\textwidth, clip=True, trim=0.2cm 0.2cm 0.4cm 0.2cm]{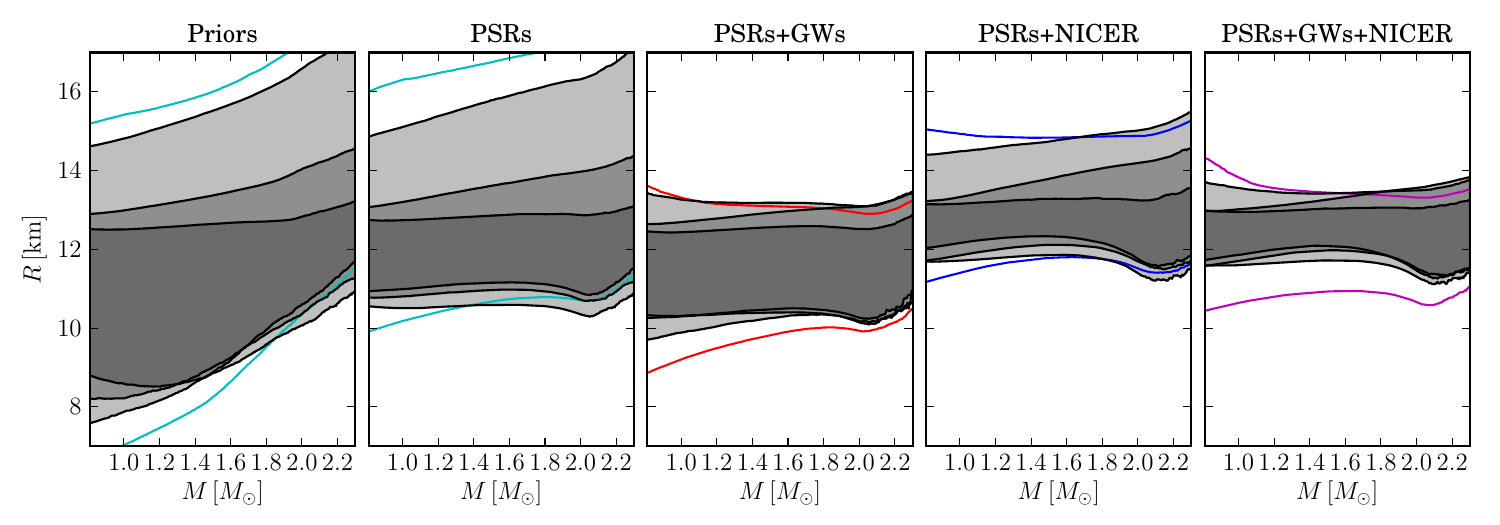}
    \caption{
        Posterior 90\% symmetric credible regions for the radius at each mass \textit{a priori} and conditioned on different data sets with the \QMC~predictions.
        Shaded regions show the posteriors conditioned on the theory predictions up to $\pmax=0.45$, $1.93$, and $10.9\,\mathrm{MeV}/\mathrm{fm}^3$ (approximately $0.5\times$, $1\times$, and $2\rhonuc$ in \QMC) and colored lines show the analogous 90\% symmetric credible regions from an agnostic analysis not conditioned on field theory predictions~\cite{Landry:2020}.
        We show the constraints obtained \textit{a priori} (far left), with only massive \PSRs~(middle left), with massive \PSRs~and \GWs~(middle), massive \PSRs~and \NICER~(middle right), and with massive \PSRs, \GWs, and \NICER~measurements (far right).
%        Compare to Fig.~\ref{fig:pressure-density}.
    }
    \label{fig:fancy EFT mr plots}
\end{figure*}

\begin{figure*}
    \includegraphics[width=\textwidth, clip=True, trim=0.2cm 1.3cm 0.4cm 0.2cm]{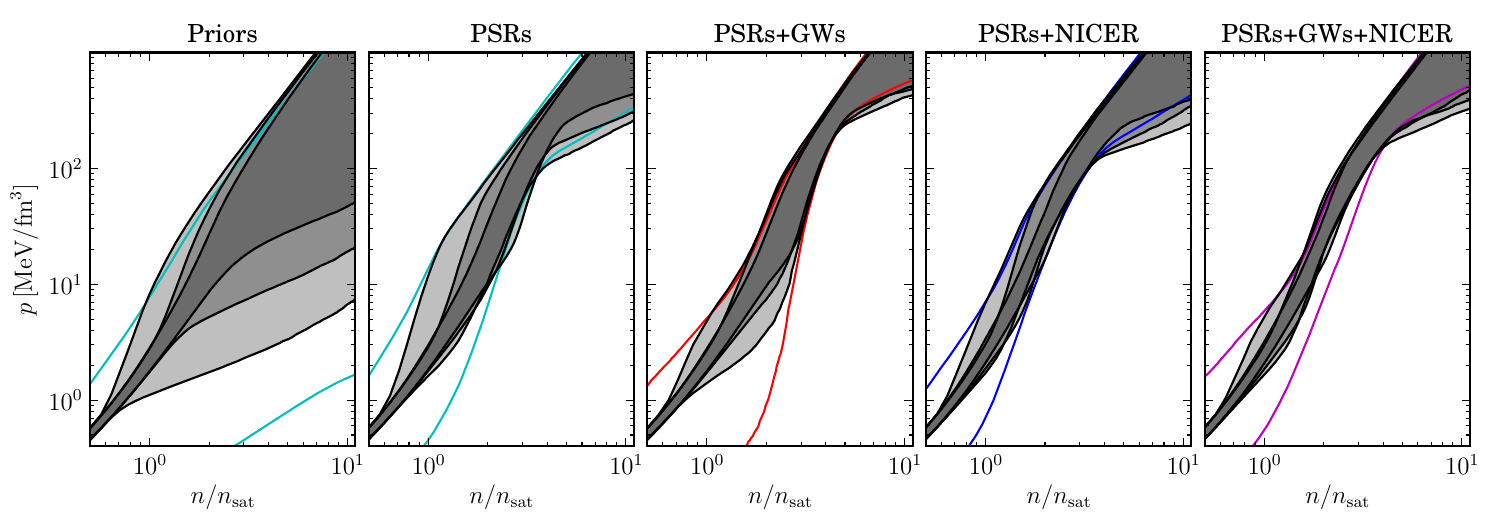}
    \includegraphics[width=\textwidth, clip=True, trim=0.2cm 1.3cm 0.4cm 0.8cm]{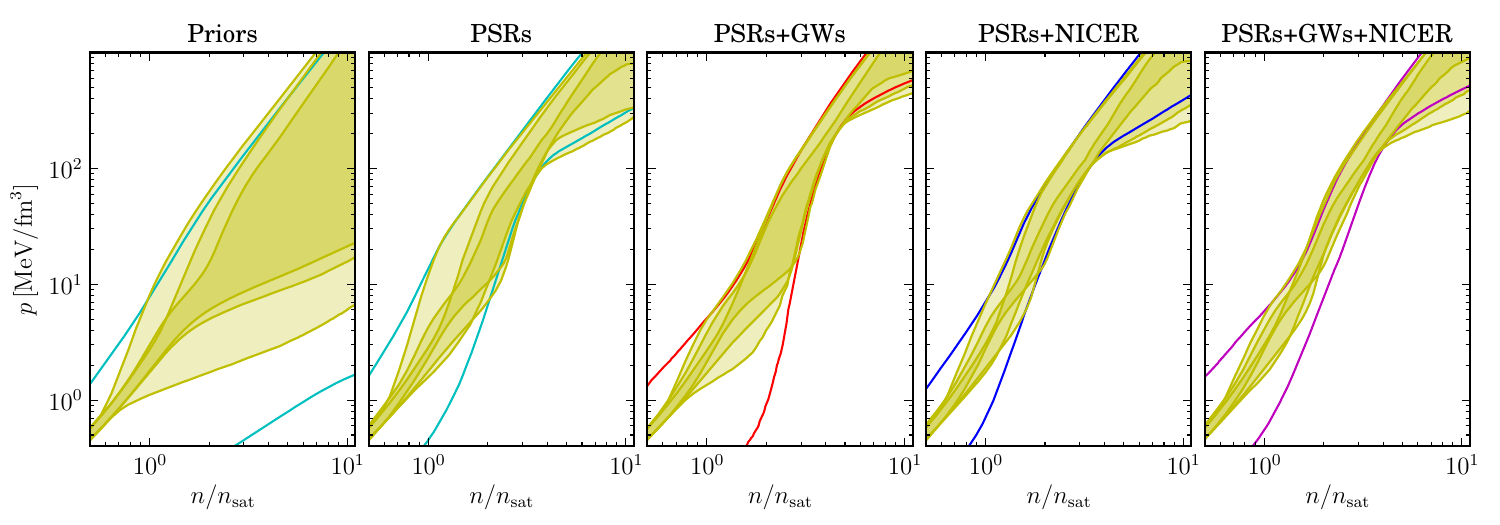}
    \includegraphics[width=\textwidth, clip=True, trim=0.2cm 0.0cm 0.4cm 0.8cm]{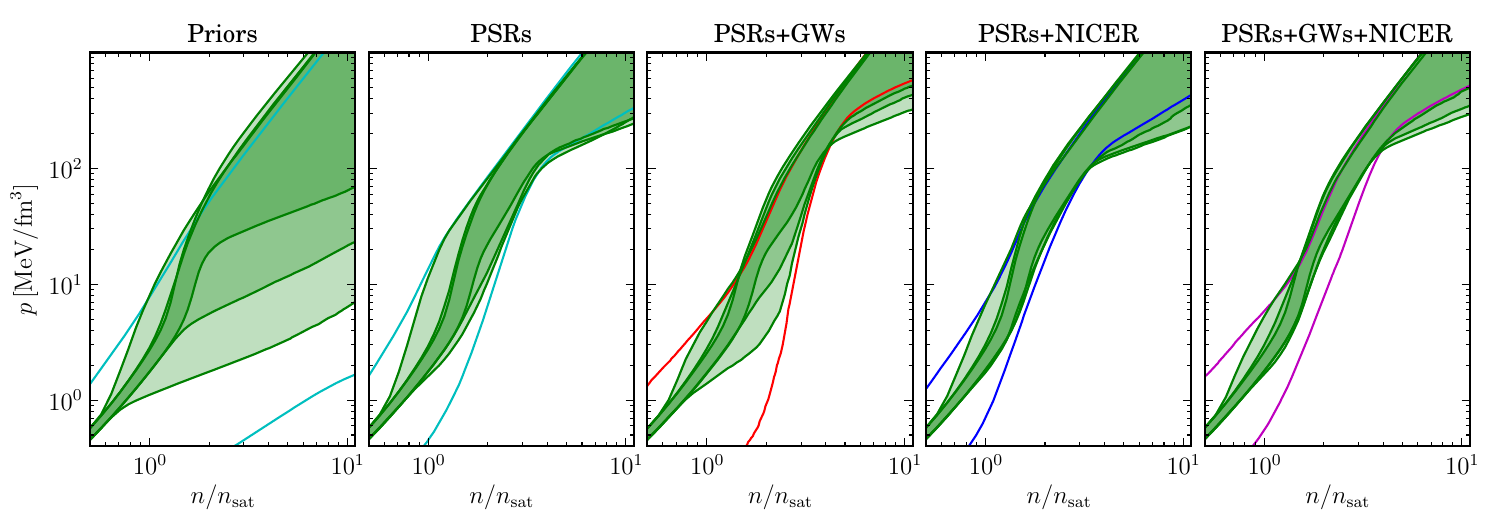}
    \caption{
        Posterior 90\% symmetric credible regions for the pressure at each density a priori and conditioned on different data sets with the \QMC~predictions (top), the soft FFT (middle), and the stiff FFT (bottom).
        Shaded regions show the posteriors conditioned on the theory predictions up to $\pmax=0.45$, $1.93$, and $10.9\,\mathrm{MeV}/\mathrm{fm}^3$ (approximately $0.5\times$, $1\times$, and $2\rhonuc$ in \QMC) and colored lines show the analogous 90\% symmetric credible regions from an agnostic analysis not conditioned on field theory predictions~\cite{Landry:2020}.
        From left to right, we show the constraints obtained \textit{a priori}, with only massive \PSRs, with massive \PSRs~and \GWs, with massive \PSRs~and \NICER~measurements, and with all astrophysical data.
        Compare to Fig.~\ref{fig:pressure-density}.
    }
    \label{fig:fancy EFT process plots}
\end{figure*}

We present a few additional figures to demonstrate how the soft and stiff FFTs compare to the real \QMC~predictions within our analysis.
Figs.~\ref{fig:fancy EFT mr plots} and~\ref{fig:fancy EFT process plots} show the posterior constraints based on different subsets of the astrophysical data in the mass-radius and pressure-density planes, respectively.
Generally, we find that \PSR~and \NICER~data favor stiffer \EOSs~and \GW~observations favor softer \EOSs.
Indeed, the lower limit on pressures is driven by a combination of the massive \PSR~and \NICER~data, while the upper limit is set primarily by \GW~observations below $\sim2\rhonuc$ and causality at higher densities, in agreement with previous studies~\cite{Landry:2020}.
This appears to be true regardless of $\pmax$.
We can also see how the agnostic nonparametric extensions above $\pmax$ tend to fill out the posterior credible regions obtained from completely agnostic analyses, as expected.

Fig.~\ref{fig:fancy EFT process plots} also shows the priors and posteriors for the soft and stiff FFTs.
In particular, it is evident that the stiff FFT significantly stiffens \textit{a priori} compared to \QMC~for $n\gtrsim\rhonuc$.
We also note that the posterior process conditioned on the soft FFT up to the maximum $\pmax$ considered is forced to the extreme upper prior bounds at low densities in order to remain consistent with the astrophysical data after it softens appreciably starting around \rhonuc.

\begin{figure*}
    \includegraphics[width=1.0\textwidth]{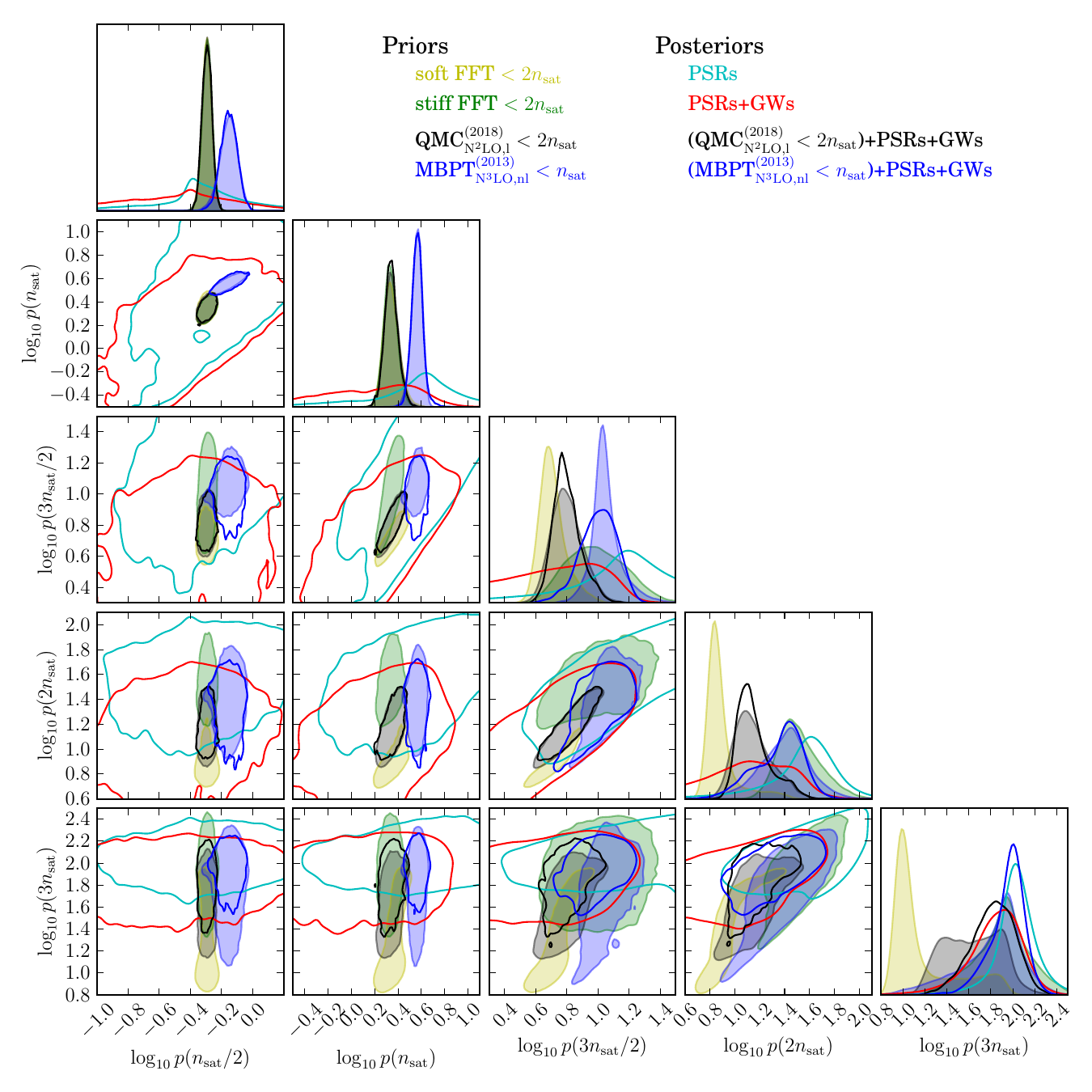}
    \caption{
        Correlations between the pressures (in $\mathrm{MeV}/\mathrm{fm}^3$) at different densities.
        Prior predictions from both the \QMC~(grey) and \MBPT~(blue) \EFT~calculations are shown in as shaded regions, and we note that the \MBPT~prior transitions to an agnostic process at lower densities than the example \QMC~prior shown here.
        We also show the soft (yellow) and stiff (green) FFT priors as shaded regions.
        Posteriors conditioned on just \PSR~data (cyan) as well as those conditioned on \PSRs~and \GWs~(red) and \PSRs, \GWs~with either \QMC~(black) or \MBPT~(blue) predictions are also shown as solid lines.
        This figure focuses on posteriors conditioned on \PSR~and \GW~observations, while Fig.~\ref{fig:corner NICER} shows the effects of \NICER~observations.
    }
    \label{fig:corner GW}
\end{figure*}

\begin{figure*}
    \includegraphics[width=1.0\textwidth]{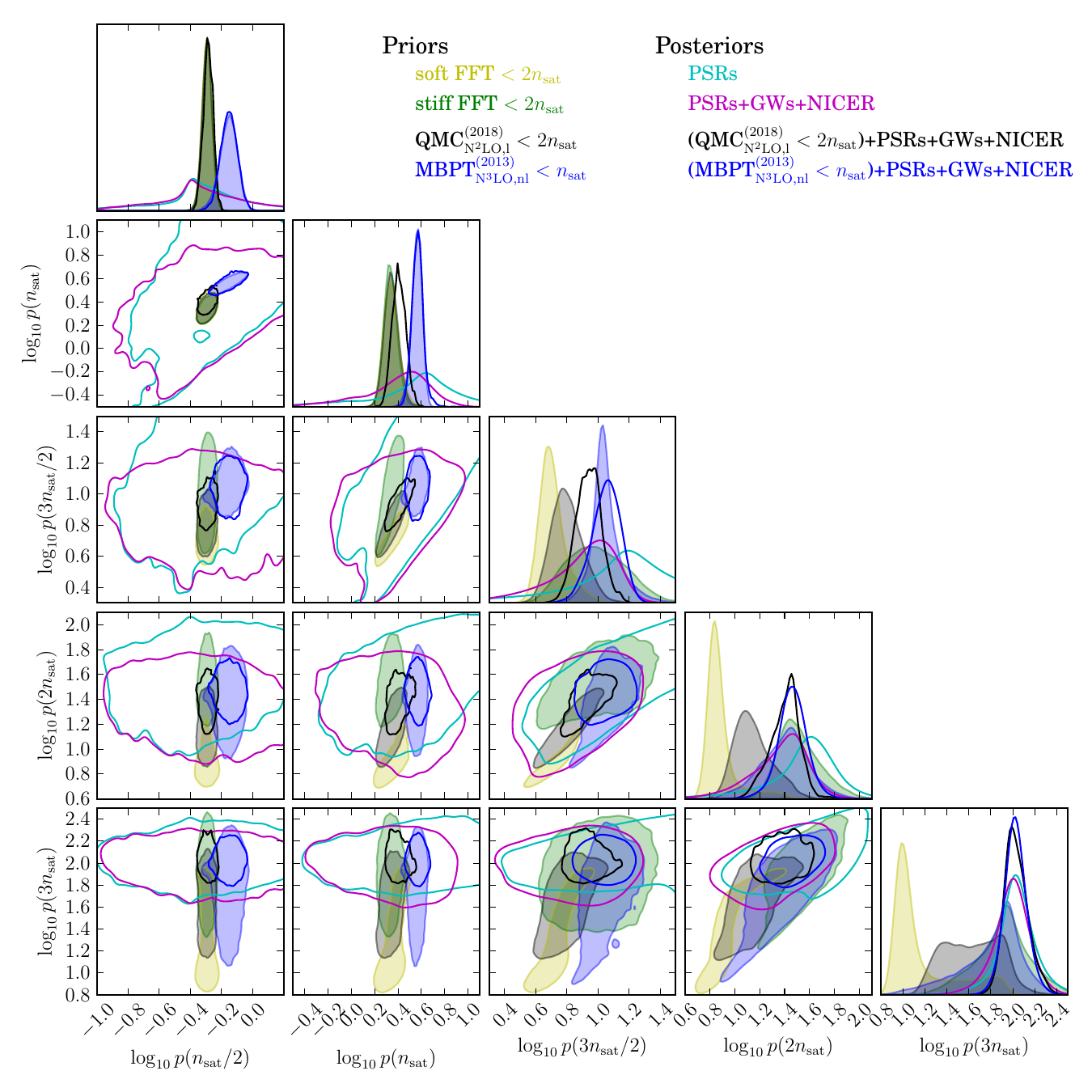}
    \caption{
        Correlations between the pressures (in $\mathrm{MeV}/\mathrm{fm}^3$) at different densities.
        This figure shows the same set of priors as Fig.~\ref{fig:corner GW}, but shows posteriors conditioned on \PSRs, \GWs, and \NICER~instead of those conditioned on only \PSRs~and \GWs.
    }
    \label{fig:corner NICER}
\end{figure*}

Figs.~\ref{fig:corner GW} and~\ref{fig:corner NICER} show correlations between the pressures at different densities both \textit{a priori} and \textit{a posteriori} for different astrophysical data sets.
Similar to Fig.~\ref{fig:probability vs pressure}, we can understand our results by comparing the prior predictions from different \EFT~calculations against the results obtained with a completely agnostic analysis that was not conditioned on \EFT~in any way~\cite{Landry:2020}.

%-------------------------------------------------
\section{Quantifying the Information}
\label{sec:information}

Fig.~\ref{fig:information} represents the amount of information obtained from various astrophysical observations when we trust \QMC~up to different $\pmax$.
Because this can be a slippery concept, we quantify exactly what we mean below.

Each of our priors is conditioned on \EFT~uncertainty up to a $\pmax$, and beyond $\pmax$ it is allowed to explore the full agnostic nonparametric prior.
To compute the evidence for each prior, we perform Monte-Carlo integrals over a finite set of realizations from that prior.
This generates a discrete set of synthetic \EOS~over which we can define distributions (as opposed to processes over functional degrees of freedom).
In this discrete set, each prior draw has equal probability (the maximum entropy distribution), and we then evaluate the marginal likelihood of each astrophysical observation for each synthetic \EOS, updating the distribution.

The Shannon information~\cite{Shannon:1948:MTCa} in base-2 ($\mathcal{I}_2$) of a probability distribution ($q$) over a discrete set of $N$ outcomes is defined as
\begin{equation}
    \mathcal{I}_2(q) = \log_2 N + \sum\limits_{i=1}^{N} q_i \log_2 q_i
\end{equation}
Consider the case where $q$ has support over only a subset of the full $N$ outcomes but each element of the subset is equally likely.
We then obtain
\begin{equation}
    \mathcal{I}_2(q) = \log_2 N - \log_2 \alpha N = - \log_2 \alpha
\end{equation}
where $0\leq \alpha \leq 1$.
We note that this is independent of the total number of outcomes initially considered, so that it does not depend on the size of our Monte-Carlo integrals.
As long the resulting posteriors do not have support for only 1 synthetic \EOS~($\alpha \gg 1/N$), the statistic should be numerically stable as we increase the size of our Monte-Carlo integrals.
We also have the natural interpretation for any $q$ that the fraction of the prior that is still viable \textit{a posteriori} is
\begin{equation}
    \alpha(q) = 2^{-\mathcal{I}_2(q)}.
\end{equation}
Fig.~\ref{fig:information} presents this quantity for the posterior distribution obtained with different astrophysical data for our priors conditioned up to different $\pmax$.
Although these statistics derived from distributions over finite sets of realizations of a process over infinitely many degrees of freedom cannot capture all the subtleties associated with the full process, and equivalent measures may not be easily produced by analyses that do not perform Monte-Carlo integrals as we do, we believe these considerations are, if nothing else, of pedagogical value.

We also note a quirk that is apparent in Fig.~\ref{fig:information}.
When we trust \EFT~up to high densities, the posterior conditioned on all astrophysical data appears to have slightly less information than the posterior conditioned on only \PSRs~and \NICER~observations.
This happens because the \GW~and \NICER~data together actually produce flatter processes (wider distributions) to some extent, particularly at high densities.
With a wider distribution, there are more viable EOS because the posterior is not as concentrated.
Physically, this is because the \GW~data likes soft things that just barely satisfy the \PSR~mass requirements.
That tends produce posteriors at high densities that are concentrated on the softest \EOS~permissible.
\NICER~data adds more weight to stiffer \EOS, and so the high density behavior is not quite as concentrated.
This is apparent in Fig. 9 of Ref.~\cite{Landry:2020}, in the first column which shows the posterior distributions for $M_\mathrm{max}$.
Nonetheless, the difference is the information reported here is relatively small, and it is clear that the \NICER~data still dominates the behavior.

%------------------------------------------------------------
\end{document}